\documentclass[lettersize,journal]{IEEEtran}

\usepackage{tikz}
\usepackage{amsmath}

\usepackage{cite}
\usepackage{amsmath,amssymb,amsfonts}
\usepackage{algorithmic}
\usepackage{graphicx}
\usepackage{booktabs}
\usepackage{textcomp}
\usepackage{multirow}
\usepackage{xcolor}
\usepackage{tabularx}
\usepackage{caption}
\usepackage[normalem]{ulem}
\usepackage{lipsum}
\usepackage{enumitem}
\usepackage{multirow}
\usepackage{makecell}
\usepackage{pifont}
\usepackage{siunitx}
\usepackage{makecell}
\newcommand{\cmark}{\ding{51}}%
\newcommand{\xmark}{\ding{55}}%
\usepackage{wrapfig}
\usepackage{enumitem}
\useunder{\uline}{\ul}{}
\def\BibTeX{{\rm B\kern-.05em{\sc i\kern-.025em b}\kern-.08em
    T\kern-.1667em\lower.7ex\hbox{E}\kern-.125emX}}

\usepackage{tabularray}
\usepackage{ragged2e}
\usepackage{balance}

\setlist{itemsep=10pt}

\usepackage{tcolorbox}

\newtcolorbox{openquestionbox}{
  colback=gray!10!white,
  colframe=gray!90!black,
  title=Open questions,
  sharp corners,
  top=1mm,
  bottom=1mm,
  left=0.5mm,  
  right=5mm,  
  boxrule=0.1mm
}

\usepackage{amsmath,amsfonts}
\usepackage{algorithmic}
\usepackage{array}
\usepackage[caption=false,font=normalsize,labelfont=sf,textfont=sf]{subfig}
\usepackage{textcomp}
\usepackage{stfloats}
\usepackage{url}
\usepackage{verbatim}
\usepackage{graphicx}
\def\BibTeX{{\rm B\kern-.05em{\sc i\kern-.025em b}\kern-.08em
    T\kern-.1667em\lower.7ex\hbox{E}\kern-.125emX}}
\usepackage{balance}
\begin{document}
\title{TinyML Security: Exploring Vulnerabilities in Resource-Constrained Machine Learning Systems}
\author{
\IEEEauthorblockN{Jacob Huckelberry\IEEEauthorrefmark{1}\IEEEauthorrefmark{3}
Yuke Zhang\IEEEauthorrefmark{2}\\
Allison Sansone\IEEEauthorrefmark{3}}
James Mickens\IEEEauthorrefmark{1}
Peter A. Beerel\IEEEauthorrefmark{2}
Vijay Janapa Reddi\IEEEauthorrefmark{1}\\

\IEEEauthorblockA{
\IEEEauthorrefmark{1}Harvard University \hspace{0.5em}
\IEEEauthorrefmark{2}University of Southern California \hspace{0.5em}
\IEEEauthorrefmark{3}Draper Laboratory
}
}
\maketitle

\begin{abstract}
Tiny Machine Learning (TinyML) systems, which enable machine learning inference on highly resource-constrained devices, are transforming edge computing but encounter unique security challenges. These devices, restricted by RAM and CPU capabilities two to three orders of magnitude smaller than conventional systems, make traditional software and hardware security solutions impractical. The physical accessibility of these devices exacerbates their susceptibility to side-channel attacks and information leakage. Additionally, TinyML models pose security risks, with weights potentially encoding sensitive data and query interfaces that can be exploited. This paper offers the first thorough survey of TinyML security threats. We present a device taxonomy that differentiates between IoT, EdgeML, and TinyML, highlighting vulnerabilities unique to TinyML. We list various attack vectors, assess their threat levels using the Common Vulnerability Scoring System, and evaluate both existing and possible defenses. Our analysis identifies where traditional security measures are adequate and where solutions tailored to TinyML are essential. Our results underscore the pressing need for specialized security solutions in TinyML to ensure robust and secure edge computing applications. We aim to inform the research community and inspire innovative approaches to protecting this rapidly evolving and critical field.
\end{abstract}

\begin{IEEEkeywords}
Tiny Machine Learning (TinyML), TinyML security, TinyML attack vectors
\end{IEEEkeywords}

\section{Introduction}
\label{sec:intro}

The computing landscape has undergone a significant transformation in recent years, driven by the proliferation of connected devices and the growing demand for real-time data processing. Although cloud computing has long been the backbone of data analysis and processing to extract intelligence, the growing need for immediate decision-making and reduced latency has precipitated a shift to edge computing~\cite{reddi}. The transition to edge computing brings the computation closer to the data source, enabling faster response times and more efficient use of network resources. Within this evolving ecosystem, the Internet of Things (IoT) has emerged as a key driver, with tens of billions of interconnected devices that improve both everyday activities and industrial operations~\cite{reddi}. 

\begin{figure}[t]
    \centering
    \includegraphics[width=0.9\columnwidth]
    {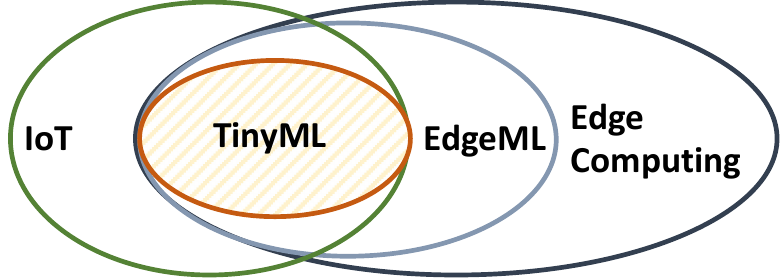}
    \caption{Venn diagram illustrating the interrelationships between IoT, Edge Computing, EdgeML, and TinyML. It illustrates how TinyML overlaps with both IoT and EdgeML within the broader scope of Edge Computing.}
    \label{fig:timeline}
\end{figure}

As these devices become more prevalent, the demand for localized and efficient computing solutions continues to grow. TinyML has emerged as a critical field at the forefront of this shift, bridging the gap between IoT devices and edge computing capabilities. Figure~\ref{fig:timeline} illustrates TinyML's unique position within the broader context of Edge Computing. By combining the resource-constrained nature of IoT devices with the computational capabilities of EdgeML, TinyML enables machine learning applications on extremely low-power devices. This convergence of technologies opens up new possibilities for intelligent and autonomous operations at the edge of networks. This innovation is set to revolutionize multiple industries, including healthcare, manufacturing, and environmental monitoring by offering localized and efficient data processing and decision-making capabilities~\cite{Abadade2023ATinyML, reddi}.

As TinyML technology swiftly progresses and integrates into a growing number of applications, a key concern frequently neglected is security. TinyML devices operate with stringent resource limitations, possessing memory and computational power often two to three orders of magnitude below that of traditional IoT or edge devices. This significant disparity in resources introduces unprecedented challenges to implementing security measures. Whereas a standard edge device might have megabytes of RAM and processors running at gigahertz speeds, TinyML devices typically function with only kilobytes of memory and processors operating at megahertz speeds. These harsh constraints render it impractical, if not impossible, to apply conventional security methods directly. Additionally, the broad deployment of these resource-limited devices in diverse and often physically accessible locations subjects them to various potential threats. Given that TinyML systems are increasingly relied upon for sensitive tasks and data, addressing these specific security challenges is imperative for their effective and responsible use.

The distinctive features of TinyML devices present a unique set of security concerns that require careful consideration. Traditional security approaches, crafted for environments with abundant resources, often clash with the stringent limitations of TinyML systems. This mismatch leads to a significant vulnerability in safeguarding these devices from various threats. First, the minimal computational power and memory of TinyML devices make it difficult to deploy strong cryptographic protocols or sophisticated authentication methods. Second, their physical presence in varied deployment settings makes them susceptible to tampering and side-channel attacks, which are especially challenging to address given the devices' constrained resources. Third, ML models, which are frequently proprietary and include sensitive information, become attractive targets for theft or manipulation. Additionally, the network-connected nature of many TinyML applications exposes them to remote attacks, while devices lack the capacity for extensive network security measures. This issue is exacerbated by the need to balance security with the functionality of these devices, ensuring that security measures do not significantly hinder their primary operations.

\begin{figure}[t]
    \centering
    \includegraphics[scale=0.6]{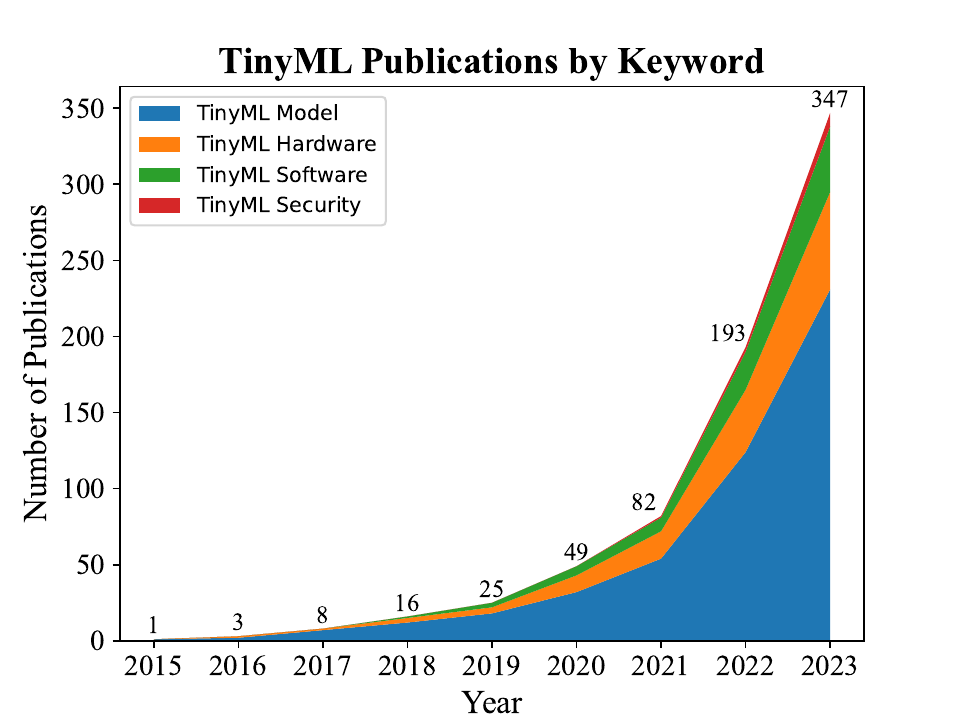}
    \caption{TinyML publication trends from 2015 to 2024. Alarmingly, few papers focus on TinyML security despite its pervasiveness. Many security-focused papers are more concerned with utilizing TinyML to enhance security rather than addressing the security of TinyML devices themselves.}
    \label{fig:trends}
\end{figure}

In this paper, we provide a thorough and organized evaluation of the security landscape within TinyML, to close the current knowledge gap by presenting a comprehensive understanding of specific security issues and countermeasures pertinent to TinyML systems. The importance of our work is emphasized by the significant gap between the rapid progress in TinyML technologies and the relatively slow pace of research on their security. Figure~\ref{fig:trends} strikingly highlights this research disparity. The figure shows the trends of TinyML research publications from 2015 to 2023. During this period, around 347 publications have been dedicated to TinyML models, hardware, and software, but only 9 have tackled the issue of TinyML security. Alarmingly, many of the security-focused papers are more concerned with utilizing TinyML to improve security than addressing the security of TinyML devices themselves. This lack of security research, despite the increasing implementation of TinyML in various fields, underscores the need for an extensive review and analysis.

To conduct our vulnerability analysis, we developed a taxonomy of edge devices, distinguishing between traditional IoT devices, EdgeML devices, and TinyML devices, thereby elucidating the distinct security challenges faced by each type. In addition, we form a detailed threat model that identifies and categorizes attack vectors and unique target artifacts for TinyML devices, laying the groundwork for future security evaluations. We assess the severity and potential impact of different attacks on TinyML systems using the Common Vulnerability Scoring System (CVSS)~\cite{CVSS}, which provides a qualitative measure for risk assessment. A crucial part of our review involves evaluating the feasibility and efficiency of conventional hardware, software, and model security techniques within the tightly restricted setting of TinyML devices. We also gather findings from prior work and offer a comprehensive perspective on both the theoretical and applied aspects of TinyML security. Finally, we demonstrate the new challenges and pinpoint promising future research avenues in safeguarding TinyML systems, taking into account the blend of machine learning and resource-limited TinyML devices. 

\textbf{Hardware Vulnerabilities:} TinyML devices face significant security challenges due to their physical accessibility and resource constraints. Side-channel attacks threaten to compromise machine learning models, while leaky interfaces, though easily exploited, can be mitigated with built-in microcontroller safeguards. Fault injection attacks pose a complex threat, capable of compromising both device and model integrity. The resource limitations of TinyML devices restrict the implementation of robust security measures, such as encrypted environments or comprehensive fault injection defenses. This landscape necessitates the development of innovative, lightweight security solutions that can effectively protect TinyML systems without overwhelming their limited computational and energy resources. Future research must focus on creating efficient countermeasures that balance security needs with the practical hardware constraints of current TinyML deployments.

\textbf{Software Vulnerabilities:} TinyML devices encounter distinct software security issues due to their limited compute, memory and network resources. Ensuring communication security, especially against eavesdropping and man-in-the-middle attacks, is crucial but challenging, as robust protocols like TLS cannot be implemented on these constrained devices. The trade-off between on-device ML inference and secure networking protocols remains largely unexamined, pointing to an important research gap. Additionally, over-the-air (OTA) updates for TinyML models pose a significant concern, as safeguarding their confidentiality, integrity, and authenticity is vital for protecting intellectual property and device functionality. Although solutions such as RIOT-ML provide some security controls, they might inadvertently increase the attack surface. Future research should concentrate on creating lightweight, standalone security mechanisms for secure communications and OTA updates, designed specifically for the resource constraints of TinyML devices. Understanding the trade-offs between ML inference, networking protocols, and security measures is essential for optimizing TinyML deployments.

\textbf{Model-specific Vulnerabilities:} TinyML systems face a range of sophisticated attacks targeting their machine learning models. Adversarial examples pose a substantial threat, especially in critical applications like healthcare and autonomous vehicles. While quantized models show similar robustness to their full-precision counterparts, the efficiency of pre-processing defenses on resource-constrained devices remains an open question. Model extraction attacks are particularly concerning, as they can facilitate more effective adversarial and model inversion attacks. Encrypting inference outputs offers a first line of defense, but the vulnerability of TinyML models compared to larger systems requires further investigation. Backdoor attacks have demonstrated effectiveness on lightweight networks, showing that they need thorough screening before deployment. The transferability of conditioned backdoor attacks across compression techniques and the feasibility of on-device defenses are areas that need exploration. Model inversion attacks pose significant privacy risks, especially for networks trained on sensitive data. While maintaining parameter confidentiality and controlling access to inference results can help, a theoretical framework for assessing model inversion attack robustness is notably absent. Future research should focus on developing resource-efficient defenses, understanding the unique vulnerabilities of TinyML models, and creating theoretical frameworks for assessing and enhancing model security in these constrained environments.

In summary, we address the critical gap in TinyML security research by providing a comprehensive analysis of the unique security challenges faced by these resource-constrained devices. Our examination covers hardware vulnerabilities, software security issues, and model-specific threats, offering insights into the complexities of protecting TinyML systems. By developing a taxonomy of edge devices, formulating a threat model, and assessing the feasibility of existing security techniques, we laid the groundwork for future advancements in TinyML security. The stark contrast between the rapid proliferation of TinyML technologies and the limited focus on their security underscores the urgency of our work. As TinyML continues to revolutionize various industries with its localized and efficient data processing capabilities, addressing these security concerns becomes paramount. This paper not only highlights the current state of TinyML security, but also identifies promising avenues for future research, paving the way for the development of innovative lightweight security solutions that can effectively protect TinyML systems without compromising their low resource efficiency and functionality.

\section{Background \& Related Work}

\renewcommand{\arraystretch}{1.5}
\label{sec:background}
\renewcommand{\arraystretch}{1.5}

\subsection{Comparison of Edge Devices}

The development of edge computing, including advances from the Internet of Things (IoT) to TinyML, has enhanced data processing capabilities at the network's edge. However, this growth has also led to novel and complex security issues. Understanding this development is key to appreciating the distinct security environment surrounding TinyML systems.

Figure~\ref{fig:timeline} showed a Venn diagram with overlaps between IoT, EdgeML, and TinyML devices, emphasizing their shared and unique attributes. Traditional {IoT} devices predominantly serve as passive data collectors, forwarding collected data to cloud servers for analysis. Although this method is effective, it usually results in significant bandwidth usage and increased latency~\cite{edge_survey}. To mitigate these challenges, the {edge computing} paradigm was introduced, bringing data processing closer to the source of collection~\cite{edge_survey}. This approach reduces network load and enhances response times, enabling the development of more advanced edge-based systems. 

Building upon edge computing, edge ML represents a significant leap forward. In this model, machine learning inference occurs directly on data collection devices such as smartphones or Raspberry Pis~\cite{TinyML_progress, micronets}. These EdgeML devices, while more resource-constrained than cloud servers, still possess considerable computational power, typically featuring gigabytes of RAM and CPU speeds in the gigahertz range.

TinyML pushes the boundaries even further by enabling machine learning on extremely resource-constrained devices, such as microcontroller units (MCUs). As shown in Table~\ref{tab:examples}, these devices operate with mere kilobytes of RAM and clock speeds in the megahertz range. The stringent resource limitations of TinyML devices are required by their deployment scenarios, which often require minimal power consumption. For example, wearable health monitors and embedded sensors in industrial plants must operate on small batteries for extended periods.

\begin{table}[t!]
\resizebox{\columnwidth}{!}{%
\begin{tabular}{|l|l|l|l|l|}
\hline
\textbf{Device} & \textbf{Class} & \textbf{Memory} & \textbf{Storage} & \textbf{Clock Freq.} \\ \hline
Raspberry Pi 5      & IoT    & 8GB   & up to 2TB & 2.4 GHz \\ \hline
Samsung Galaxy S24  & EdgeML & 8GB   & 512GB     & 3.2 GHz \\ \hline
NUCLEO-L4R5ZI       & TinyML & 640KB & 2MB       & 120 MHz \\ \hline
CY8CPROTO-062-4343W & TinyML & 1MB   & 2MB       & 150 MHz \\ \hline
DISCO-F746NG        & TinyML & 340KB & 1MB       & 216 MHz \\ \hline
\end{tabular}%
}
\caption{Exemplar devices for the IoT, EdgeML, and TinyML paradigms, drawing information from the industry-standard MLPerf benchmarks 
~\cite{mlperf_inf}~\cite{mlperf_tiny}.}
\label{tab:examples}
\end{table} 

\subsection{Research Gaps in Related Work}

TinyML presents distinct security challenges. Although EdgeML devices with ample resources can adopt security strategies similar to those in servers and laptops, TinyML devices are significantly constrained. Their limited computational power renders many conventional security measures unfeasible or impractical to combine. This scarcity of resources shapes a unique security scenario for TinyML, which requires novel threat mitigation strategies. As we explore TinyML's security aspects more deeply, it becomes evident that the field needs innovative solutions that cater to its specific limitations.

We focus on a thorough examination of potential security risks and their resolutions in the TinyML realm. A detailed examination of the existing literature, summarized in Table~\ref{tab:works}, indicates that previous reviews have thoroughly addressed several important aspects of the TinyML field---with the exception of security. The prior art covers TinyML models~\cite{emerging_tech, sota_prospects, Abadade2023ATinyML, ML_oriented}, software frameworks~\cite{reddi, emerging_tech, sota_prospects, Abadade2023ATinyML, ML_oriented, tools_apps}, hardware frameworks~\cite{reddi, emerging_tech, sota_prospects, ML_oriented}, and applications~\cite{reddi, emerging_tech, sota_prospects, Abadade2023ATinyML, ML_oriented, tools_apps}. 

\begin{table}
\vspace{7pt}
\centering
\begin{tabular}{lccccccccc}
\toprule \textbf{Paper Title}
& \multicolumn{7}{c}{\textbf{Characteristics}} \\
\cmidrule(lr){2-9}
& \rotatebox{90}{\scriptsize TinyML Models}
& \rotatebox{90}{\scriptsize TinyML Software}
& \rotatebox{90}{\scriptsize TinyML Hardware}
& \rotatebox{90}{\scriptsize TinyML Applications}
& \rotatebox{90}{\scriptsize TinyML Threat Modeling}
& \rotatebox{90}{\scriptsize Edge Devices Taxonomy}
& \rotatebox{90}{\scriptsize Security Measures for TinyML}
& \rotatebox{90}{\scriptsize Future Security Research Leads} \\
\midrule
Current Progress~\cite{reddi} & \xmark  & \cmark & \cmark & \cmark &  \xmark & \xmark & \xmark & \xmark \\
Emerging Technology~\cite{emerging_tech} & \cmark  & \cmark & \cmark & \cmark &   \xmark & \xmark & \xmark & \xmark \\
SOTA~\cite{sota_prospects} & \cmark  & \cmark & \cmark & \cmark &  \xmark & \xmark & \xmark & \xmark \\
Comprehensive Survey~\cite{Abadade2023ATinyML} & \cmark  & \cmark & \xmark & \cmark &  \xmark & \xmark & \xmark & \xmark \\
Future Research~\cite{tools_apps} & \xmark  & \cmark & \xmark & \cmark & \xmark & \xmark & \xmark & \xmark \\
ML Oriented Survey~\cite{ML_oriented}& \cmark & \cmark & \cmark & \cmark &  \xmark & \xmark & \xmark & \xmark\\
\midrule
\textbf{This work} & \cmark  & \cmark & \cmark & \cmark &  \cmark & \cmark & \cmark & \cmark \\
\bottomrule
\end{tabular}
\caption{Comparison of the contributions of our TinyML review and those of others. Other review and survey papers focus largely on topics such as TinyML models, frameworks, and applications. Conversely, our paper does a critical examination of the security aspects of these devices.}
\label{tab:works}
\end{table}

Prior work does not adequately address the deficiencies in addressing TinyML's security dimensions. Specifically, these works do not adequately cover threat modeling for TinyML, security protocols tailored to TinyML settings, or future research paths in security. Furthermore, there is a clear lack of a detailed classification of edge devices with regard to TinyML security. This gap in security-oriented analysis highlights the need for a focused examination of TinyML security.

In contrast, our work addresses critical gaps in security analysis within TinyML. We extend beyond prior work by presenting a comprehensive threat model for TinyML systems, introducing a novel taxonomy of edge devices that considers their security implications, exploring specific security measures tailored for TinyML environments, and outlining future research directions in TinyML security. By addressing these previously unexplored areas, our review offers valuable insight into the evolving landscape of TinyML security. This approach not only complements the existing literature but also provides a foundation for future research on securing TinyML systems. 

\section{TinyML Threat Overview}
\label{sec:attack-surface}

\begin{figure*}[htbp]
    \centering
    \includegraphics[width=\textwidth]{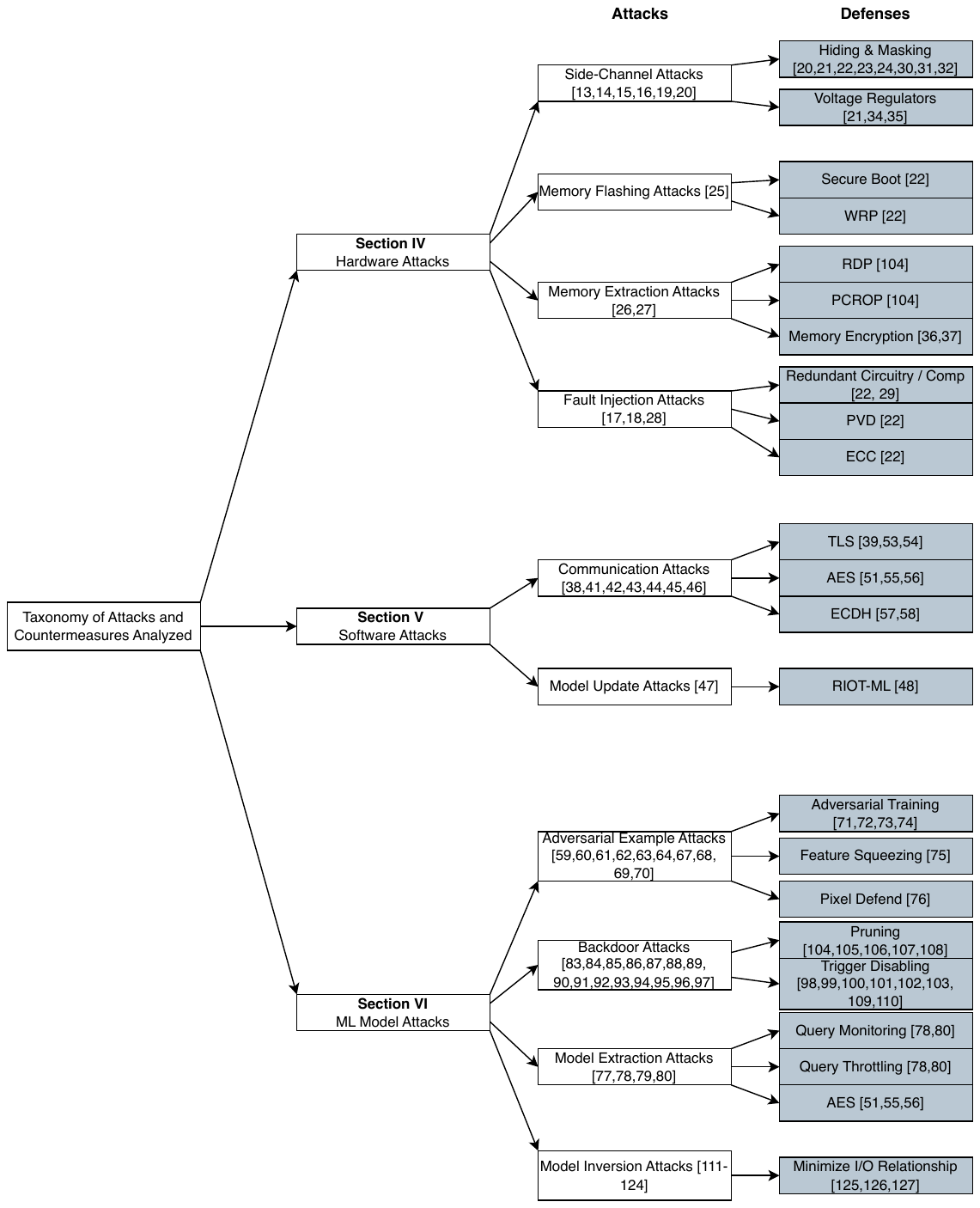}
    \caption{Taxonomy of attacks and countermeasures covered in this paper. In each section, we cover several attacks associated with the hardware, software, and model components of TinyML devices. For each of these attacks, we critically examine the feasibility of relevant countermeasures for TinyML class devices.}
    \label{fig:full-page}
\end{figure*}

\begin{figure*}[htbp]
    \centering
    \includegraphics[width=1\textwidth]{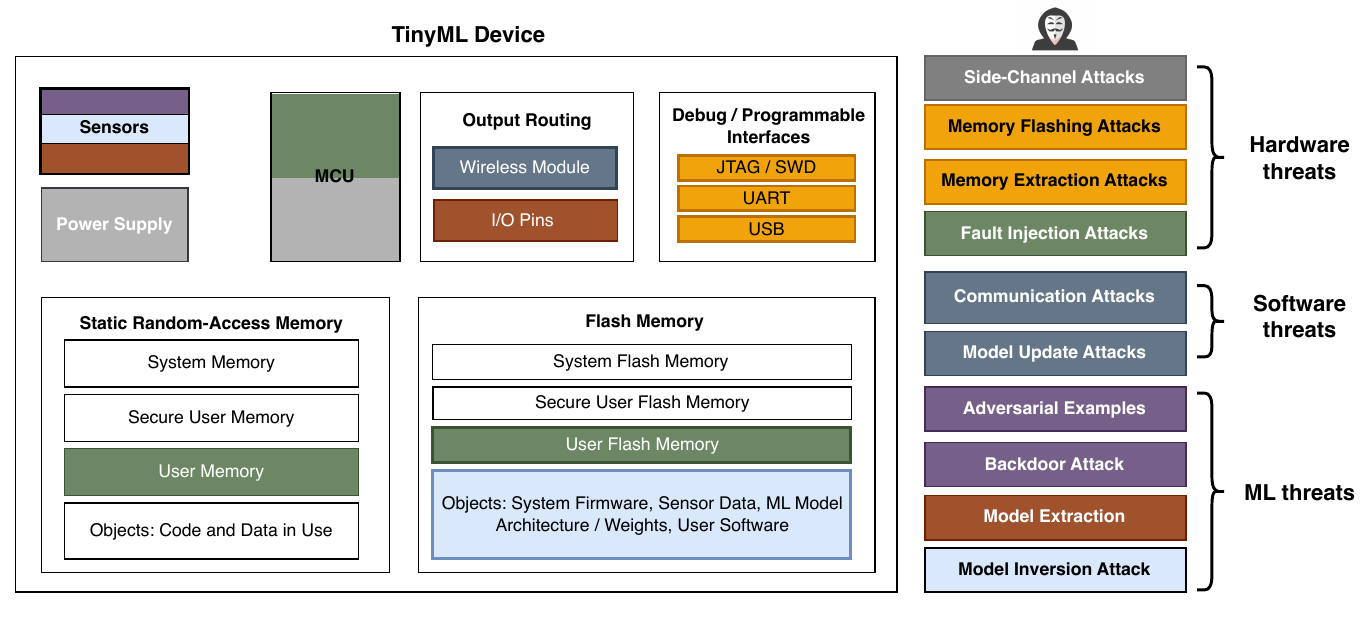}
    \caption{Low-Level TinyML Threat Model: Colors indicate the correspondence between attacks and their respective attack surfaces. Threats are categorized into hardware threats, software threats, and ML threats.}
    \label{fig:threat_model}
\end{figure*}

It is impossible to provide an exhaustive overview of the security threats associated with any system. As such, we create our threat model based on the high-level diagram of an idealized TinyML device depicted in Figure \ref{fig:threat_model}. In this figure, the color map indicates potential entry points for each attack. For our analysis, we assume that an attacker is able to access the device either \textit{physically} or \textit{remotely}. We also assume that the attacker is able to generate arbitrary packets in the device's logical network and that they have the tools and capabilities to carry out complex physical attacks.

We select attacks largely based on their plausibility given these assumptions, their coverage in research, and their relevance in practical scenarios. In addition, we choose traditional classes of attacks that present novel considerations in the context of TinyML, and attacks whose countermeasures usually require high computational overhead, making them infeasible on TinyML class devices in some cases. Specifically, several recent works have shown new side-channel attacks targeting embedded ML models~\cite{fpga_sema},~\cite{FPGA_DPA},~\cite{leaky_nets},~\cite{CSI_NN}. In addition, side-channel defenses often come at the cost of high computational overhead. There are increased privacy and security considerations for memory flashing and memory extraction attacks (leaky interface attacks), and fault injection attacks with the introduction of ML models on resource-constrained edge devices~\cite{FIA},~\cite{bit_flip}. The presence of model update mechanisms on these devices and the usage of lightweight communication protocols introduce several unique considerations for confidentiality and TinyML device integrity. 

Finally, the adversarial robustness of TinyML models is an active area of research, where not much is known about the effects of compression methods on model robustness. As this paper aims to do a system-level security analysis of TinyML devices, we organize the attack vectors into hardware, software, and machine learning security challenges. We do not cover niche exploits or attacks targeting trusted execution environments (TEEs) (e.g., Arm TrustZone) in this paper as many TinyML class devices do not support them. An overview of our selected attacks and analyzed countermeasures can be found in Figure \ref{fig:full-page}. In this figure, we organize our selected attacks into hardware, software, and ML model attacks. This structure is reflected in the following sections of this paper. We also list relevant countermeasures that we analyze and can be used to defend against each respective attack.

\begin{table}[t]
\centering
\small
\begin{tabular}{|c|c|c|c|c|c|c|c|}
\hline
\textbf{Attack} & \textbf{Sev.} & \textbf{AV} & \textbf{AC} & \textbf{PR} & \textbf{UI} & \textbf{S} & \textbf{CIA} \\ \hline
SCA & M (4.2) & P & H & N & N & U & H/N/N \\ \hline
Memory Ext. & M (4.6) & P & L & N & N & U & H/N/N \\ \hline
Memory Fla. & M (6.1) & P & L & N & N & U & N/H/H \\ \hline
FIA & M (5.7) & P & H & N & N & U & N/H/H \\ \hline
Eavesdr.. & M (6.5) & A & L & N & N & U & H/N/N \\ \hline
MITM & H (7.5) & A & H & N & N & U & H/H/N \\ \hline
Model Upd. & H (7.5) & A & H & N & N & U & H/H/H \\ \hline
Adv. Ex. & M (4.2) & P & H & N & N & U & N/H/N \\ \hline
Model Extr. & M (4.2) & P & H & N & N & U & H/N/N \\ \hline
Backdoor & M (4.2) & P & H & N & N & U & N/H/N \\ \hline
Model Inv. & M (4.2) & P & H & N & N & U & H/N/N \\ \hline
\end{tabular}
\caption{CVSS Description of Attacks Covered. The CVSS scores are presented in the order in which we cover the attacks in this paper, beginning with hardware attacks, followed by software attacks, and concluding with attacks on the ML model. In most cases, we assume that the attacks are conducted through physical media.}
\label{tab:CVSS}
\end{table}

We provide a CVSS rating and vector string to create a common understanding and quantify the severity of each attack in our threat model~\cite{CVSS}. A CVSS score is an industry standard for providing a qualitative metric for the severity of an attack if carried out. The score is placed on a scale from one to ten, where 0-3.9 is low severity, 4.0-6.9 is medium, and 7.0-10.0 is high. A low rating corresponds to limited adverse effects on a system, medium corresponds to serious adverse effects, and high corresponds to catastrophic adverse effects. These scores are calculated based on six different inputs that compose the vector string: attack vector (AV), attack complexity (AC), privileges required (PR), user interaction (UI), scope (S), confidentiality (C), integrity (I), availability (A). More can be read about these inputs and how the scores are calculated in the CVSS specification document created by First ~\cite{CVSS}. We provide CVSS scores in Table \ref{tab:CVSS} for each attack we cover in the following sections of the paper.

\section{TinyML Hardware Security Challenges}
\label{sec:hardware-challenges}

\begin{figure*}
    \centering
    \includegraphics[width=\textwidth]{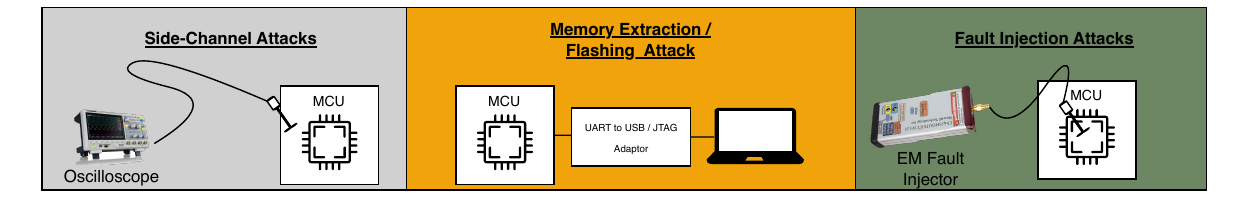}
    \caption{TinyML Hardware Attack Vectors. The primary attacks of interest are SCA, Leaky Interface Attacks, and FIAs. In this figure, very high level descriptions of how these attacks are carried out are shown.}
    \label{fig:hardware_attacks}
\end{figure*}

Given the overview in Section ~\ref{sec:attack-surface}, the hardware attacks of concern in TinyML deployments are side-channel attacks, leaky interface attacks, and fault injection attacks. Each subsection will provide a detailed analysis of the attack methods and outline specific strategies to mitigate these threats. After these descriptions, we explore the viability of the enumerated controls for TinyML implementations. A high-level overview of the hardware attacks we cover can be found in Figure \ref{fig:hardware_attacks}.

\subsection{Side-Channel Attacks}
\label{subsec:sca}
\textbf{Overview:} \textit{Side-Channel Attacks} (SCA) are a broad and well-studied class of attacks on embedded systems. These attacks hinge on taking advantage of signals a device emits during computation that leak information~\cite{phys_SCA}. Some side-channels of note are timing, power, and EM~\cite{phys_SCA}. For the purposes of this paper, we will focus on power and EM side-channels as they are the most relevant to embedded systems. The primary goal of side-channel attacks is to steal secret encryption keys in traditional embedded and IoT settings~\cite{phys_SCA}. However, in the context of TinyML, more recent research has shown promise in EM and Power Analysis Attacks for stealing machine learning models~\cite{phys_SCA}. In addition, many SCA defenses have high costs in terms of die area and computational overhead, possibly making them infeasible to implement in TinyML use cases.

\textbf{Attack Description:} In a power analysis attack, the attacker probes the power supply to detect changes in power consumption. There are two traditional kinds of power analysis attacks: simple power analysis (SPA) and differential power analysis (DPA). In a SPA attack, an attacker will visually inspect power consumption traces to identify the power consumption patterns associated with different computational operations. In these cases, the attacker usually only has access to a small sample of power traces that were produced by a small number of inputs and outputs. DPA attacks are more robust and more likely to be carried out on TinyML devices due to their physical accessibility. In a DPA attack, the attacker uses statistical methods to derive secrets with more confidence and a higher probability of success~\cite{power_analysis}. These kinds of attacks are possible when the attacker has access to numerous power traces produced by thousands of inputs and outputs. Similar to power analysis attacks are EM analysis attacks. The nomenclature is shared between these attacks where simple EM analysis (SEMA) attacks rely on visual inspection of traces while differential EM analysis (DEMA) attacks rely on statistical methods~\cite{phys_SCA}.

\textbf{Considerations for TinyML:} Batina et al.~\cite{CSI_NN} demonstrated a proof-of-concept exploit leveraging SEMA and correlational EM analysis (CEMA) to reverse engineer both a multi-layer perceptron (MLP) and convolutional neural network (CNN) on ARM Cortex-M3 and Atmel ATmega328P microcontrollers. In their results, they were able to recover the activation functions, number of layers and neurons, the number of classes, and weights of the targeted neural network~\cite{CSI_NN}. In Leaky Nets~\cite{leaky_nets}, the authors were able to achieve similar results on several MCU platforms using SPA rather than SEMA and CEMA although they primarily aimed to recover only network weights and biases. Two more studies were shown to provide similar results on FPGA platforms rather than MCUs in~\cite{fpga_sema}~\cite{FPGA_DPA}. The amount of success researchers have had in exploiting these SCAs on TinyML platforms poses a large confidentiality and integrity threat for these devices. If not mitigated, attackers can steal intellectual property and pivot into other attacks such as \textit{adversarial example attacks}. Given this description, our inputs to the CVSS calculator are a physical attack vector, high attack complexity, no privileges required, no user interaction required, an unchanged scope, and high confidentiality risk. These inputs result in a medium score of 4.2 on the CVSS scale.

\textbf{Countermeasure Overview:} There are several traditional security controls to prevent power analysis and EM analysis attacks however, the most common approaches are hiding and masking~\cite{EM_def}. Hiding involves changing the device's power consumption and EM emission such that there is no longer a correlation between different types of computation and emissions. This can be achieved through introducing randomness into the power consumption or by making power consumption uniform across all operations. Masking is a method that also may be implemented in many different ways that involves masking intermediate values that are computed over with random values, making traces unpredictable. These may be implemented in cryptographic libraries that feature options such as delays or fake instructions~\cite{STM32MCUsSecurity2024}. While these are the most common defenses against these SCA, both of these countermeasures require significant computational overhead, meaning they may not be plausible for TinyML use cases. To read more about how these controls may be implemented, see~\cite{power_analysis_book},~\cite{power_analysis_dis},~\cite{power_analysis}. In the context of TinyML, since many model recovery studies hinge on making use of signatures from activation functions, researchers have also proposed constant time implementations for these~\cite{CSI_NN}. However, again, constant time activation implementations have significant computational costs~\cite{CSI_NN}. Outside of masking and hiding, some more efficient, although perhaps less effective, controls for these attacks have been proposed, such as using voltage regulators, random voltage dithering, and shielding~\cite{EM_def}.

\begin{openquestionbox}
\begin{itemize}
\it 
    \item To what extent can robust hardware implementations of hiding and masking techniques be developed to meet the area constraints of TinyML devices?
    \item What are the most effective strategies for implementing area-efficient, constant-time activation functions in hardware for machine learning applications?
    \item How can hardware-based countermeasures against side-channel attacks be integrated into commodity microcontrollers given the fragmented ecosystem?

\end{itemize}
\end{openquestionbox}

\subsection{Leaky Interface Attacks}
\label{subsec:leaky}
\textbf{Overview:} Leaky interfaces are communication channels on embedded devices that have use cases such as debugging, transmitting outputs, or programming the device but expose more information or access than intended~\cite{MLSys_book}. Examples of these interfaces include JTAG, UART, or USB. If access to these interfaces is not properly configured to be secure, it could allow an attacker to arbitrarily write or read data to and from the device. If present in TinyML devices, these vulnerabilities could allow attackers to arbitrarily read and write the embedded ML model, giving them several ways to pivot into other attacks and affect device confidentiality, integrity, and availability. TinyML class microcontrollers usually have some built defenses for these attacks, however, they may not be as robust as the defenses offered on full-scale devices. As reading or writing data requires differing techniques and interface access, we will split this section into \textit{Memory Extraction Attacks} and \textit{Memory Flashing Attacks}, where extraction allows an attacker to read data and flashing allows an attacker to write.

\subsubsection{Memory Extraction Attacks}
\label{subsec:extract}
\textbf{Attack Description:} Memory extraction attacks are usually \textit{very} easy to exploit in hobbyist implementations or in commercial applications where security is an afterthought. Research by Vasile et al. shows how these attacks were able to be executed on three separate products from Amazon and several other devices from tech giants such as Google, LG, Samsung, and Asus~\cite{firm_extract}. These attacks are made possible by debugging interfaces such as UART, JTAG, and SWD, or via a raw flash dump~\cite{firm_extract}. Incorrect management of debugging interfaces allows attackers to gain easy access to device memory as UART connections alone are often enough to give the attacker access to a shell on the device, while with the correct device programmer, a JTAG/SWD connection allows the attacker to retrieve a full device memory dump. While more nuanced, raw flash dumps can be accomplished easily on eMMC storage devices with the help of an SD card reader. For step-by-step proof of concept exploits for these interfaces, see~\cite{firm_extract},~\cite{blackhat}.

\textbf{Considerations for TinyML:} The device firmware and data collected by edge devices are both critical pieces of information that must be protected with robust security measures. Unauthorized access to either of these elements can have severe consequences. If attackers gain access to the device firmware, they can examine the code base for vulnerabilities, enabling them to find places to pivot and further exploit the device. Furthermore, in the context of TinyML, this could leak intellectual property such as model weights, allow the attacker to reverse engineer the ML model, and pivot into \textit{Adversarial Example Attacks}. Even more concerning, attackers could access data collected by onboard sensors, leading to critical privacy breaches. The inputs to CVSS for this attack are a physical attack vector, low attack complexity, no privileges required, no user interaction, unchanged scope, and high confidentiality risk, giving it a medium CVSS score of 4.6.\vspace{-\parskip}

\textbf{Countermeasure Overview:} Fortunately these attacks are easy to mitigate with proper care and hardware support from your device. A naive solution to preventing access to these debugging interfaces is by physically scraping them off the device. However, with the proper equipment, an attacker can reinstall these components and probe into the device as if they were never removed. The proper controls for these interfaces will be device dependent but should be similar to what STM has implemented in their Readout protection (RDP) and Execute-only firmware (PCROP) security features ~\cite{STM32MCUsSecurity2024}. These are options a user can set such that it makes it impossible for attackers to access device firmware through debugging interfaces. More specifically, at its highest level of protection, RDP prevents any accesses to flash memory, SRAM, and RDP options registers by debugging interfaces while still allowing firmware updates~\cite{STM32MCUsSecurity2024}. PCROP takes this protection a step further, allowing the developer to store selected firmware in a memory region that may \textit{only} be fetched by the CPU instruction bus~\cite{STM32MCUsSecurity2024}. For added layers of protection, it is also recommended to encrypt sensitive pieces of firmware or data at rest and flush the data from the device once it has served its purpose in the device pipeline. These implementation decisions make it so that even if the attacker finds a way around these controls they cannot decipher the sensitive content and are less likely to find meaningful data in the first place. Unfortunately, these final two controls are the only controls for attacks on eMMC memory devices other than to use boards that have memory units that are more difficult to obtain raw dumps from.

\subsubsection{Memory Flashing Attacks}
\label{subsubsec:flash}
\textbf{Attack Description:} Related to extraction attacks but more active are memory flashing attacks. In these attacks, a malicious actor is able to write, rather than read, arbitrary firmware onto a vulnerable device. The presence of these vulnerabilities on TinyML devices could allow the attacker to arbitrarily modify the embedded ML model, preprocessing code, and the underlying firmware, leading to almost arbitrarily bad consequences. Luckily, the countermeasures for these attacks are relatively straightforward and accessible even on resource-constrained devices.

\textbf{Considerations for TinyML:} The heart of TinyML devices is the embedded ML model, making this the key target for an attacker to compromise device integrity and availability. With a flashing and extraction exploit, the attacker could reach ends such as creating a pivot point in the preprocessing code for \textit{adversarial example attacks}, conducting a \textit{backdoor attack} by modifying model weights in place, replacing the entire legitimate model with their own, or installing a backdoor to give them access to the model update framework. Without extraction vulnerabilities, the attacker could likely achieve similar ends in the worst case and could only replace all of the firmware with malicious code in the best case. In a grossly misconfigured device, an attacker could easily overwrite the device firmware via a USB or debugging interface connection. Our inputs to the CVSS calculator for this attack are physical attack vector, low attack complexity, no privileges required, no user interaction, an unchanged scope, and high integrity and availability risks, giving the attack a CVSS score of 6.1. 

\textbf{Countermeasure Overview:} Two ways to effectively defend against these kinds of attacks are through using secure boot and the additional hardware controls available on robust microcontrollers. Secure boot is a feature that establishes a hardware root of trust by authenticating the device firmware before the device is able to boot. This is often achieved by having a manufacturer's public key burned into the device hardware at the time of manufacture which is used to verify the integrity and authenticity of device firmware using cryptographic signatures and hashing. With this control, the device can prevent unauthorized or malicious firmware from being loaded, thereby protecting the system from tampering and unauthorized modifications. In addition to the manufacturer root key used to verify OEM firmware, auxiliary keys are often also burned into the firmware that may be used for developer use cases. These auxiliary keys can be used to encrypt the application firmware along with developer keys used to sign and authenticate the application firmware they flash onto the device. In addition to these controls some manufacturers, like STM, provide additional controls such as their write protection (WRP). WRP allows developers to mark specific memory regions that may not be written to or erased~\cite{STM32MCUsSecurity2024}. Using these controls, the device could only be compromised if the keys used for these methods were somehow exposed.

\begin{openquestionbox}
\begin{itemize}
\it 
\item How can hardware or software countermeasures be effectively implemented to harden leaky interfaces in TinyML class devices, given their lack of operating systems and memory isolation schemas?
\item To what extent can tamper detection countermeasures be implemented in TinyML devices to identify unauthorized physical access or manipulation? 
\item What are the optimal response strategies to detected tampering events, considering options such as data wiping, power-off protocols, or system resets?
\end{itemize}
\end{openquestionbox}

\subsection{Fault Injection Attacks}
\textbf{Overview:} Fault injection attacks (FIA) are a broad class of attacks that can be used to reach several ends. The traditional use case for these attacks is to attempt to steal cryptographic keys, similar to SCA. More recent work has shown FIAs for use cases such as bypassing secure boot and, in the context of TinyML, faulting ML models~\cite{FIA},~\cite{bit_flip}. Due to the diversity of this class of attacks, it is often difficult and expensive to implement comprehensive countermeasures for FIAs.

\textbf{Attack Description:} FIAs make use of several \textit{fault models}. These fault models determine what behavior the attacker hopes to induce on the target device. Examples of fault models include bit flips, bit set/reset, random byte, instruction skip, execution faults, or stuck-at faults~\cite{FIA}. Just as this model appears, it allows an attacker to entirely skip over an instruction that is being executed. Instruction corruption is similar to this fault model and can be used to bypass secure boot~\cite{secure_boot_bypass}. For more information on each of these fault models, see~\cite{FIA}. Fault models can be delivered in several ways including clock/voltage glitching, optical fault injection, and EM fault injection. Clock and voltage glitching attacks often make use of a MCU or FPGA to manipulate the internal clock or power supply of a device. Optical FIAs are diverse and can make use of almost any tool that releases a concentrated light source. The most popular tools are lasers as they are they offer high precision and attack reproducibility~\cite{FIA}. Of course, EM fault injections use devices such as EM pulse generators to throw EM at the targeted device. Experimental setups for these delivery methods and some more analysis can be read about in~\cite{FIA}. 

\textbf{Considerations for TinyML:} Perhaps the most interesting fault model for TinyML is the instruction skip model as it can be used for privilege escalation, key extractions, and neural network misclassification ~\cite{FIA}. Outside of instruction skips, the bit-flip model has also been shown to degrade the performance of CNNs by flipping the bits of model weights stored in DRAM~\cite{bit_flip}. While these kinds of attacks have the potential to do catastrophic damage to devices, they are often costly, require extremely low-level knowledge of the target device and nation-state level tooling, and are highly invasive (meaning they can often break the device). This attack is a physical attack vector, has high complexity, requires no privileges, no user interaction, has an unchanged scope, and has high integrity and availability risks. As such, these attacks receive a medium score of 5.7 on the CVSS scale.

\textbf{Countermeasure Overview:} Unfortunately there is not much the developers themselves can do to defend their TinyML devices against FIAs other than by choosing to use a robust microcontroller. Additionally, due to the diversity of FIAs, even robust microcontrollers have difficulty comprehensively defending themselves against these kinds of attacks. Many controls for FIAs are implemented at the hardware level by doing things such as implementing redundant circuitry, clock-check blocks, voltage and frequency limit checking, or sensor-based fault detection~\cite{fia_counter}. Software controls are computationally expensive as they include methods such as redundant computation~\cite{fia_counter}. Other methods may be tedious to implement and not as robust, such as using strict comparisons when branching or using values that are incremented by prime numbers within branches and ensuring they are the number that was expected~\cite{STM32MCUsSecurity2024}. Some more realistic ways to defend against FIAs are by using MCUs that have error code correction mechanisms (ECC), clock security mechanisms, voltage regulators, and physical shielding. Again, STM provides a set of robust chips that include many of these security features. Firstly, they include anti-tamper mechanisms that continuously observe the MCUs voltage to defend against voltage glitching delivery mechanism. This mechanism is referred to in their documentation as their programmable voltage detector (PVD). They also provide ECC to detect up to 2 bits of error and up to 1 bit of error correction. This can help defend against bit set/reset and bit flip fault models~\cite{STM32MCUsSecurity2024}. 

\begin{openquestionbox}
\begin{itemize}
\it 
\item To what extent can power- and die area-efficient redundant computations and integrity checks be implemented on TinyML class devices without compromising their performance or size constraints?

\item What generalizable FIA countermeasures can be identified against multiple fault models and can be efficiently implemented for TinyML devices?
\end{itemize}
\end{openquestionbox}

\subsection{Hardware Countermeasure Viability for TinyML}

Throughout this section, we have covered several countermeasures that have been proven to effectively defend against the attacks covered. However, many of these well-studied countermeasures may not be feasible to implement in the context of TinyML due to resource constraints or do not offer very much robustness to the outlined attacks. Here, we will analyze each proposed countermeasure to determine the feasibility and robustness for TinyML class devices. In this section and the viability sections following, our enumerate lists will pair countermeasures to attacks.

\begin{enumerate}[noitemsep]
    \item \textbf{Hiding, Masking, Constant Time Activation Functions (SCA):} These countermeasures have been proven to be some of the most effective methods to thwart power and EM analysis attacks, however, are likely not feasible to implement on TinyML class devices because of increased power, area, and computational costs~\cite{EM_def, mcu_sca_def, hiding_mask_cost,rekey}. It has been shown that just first-order S-Box masking for AES doubles computation time or requires more than 64KB of memory to store pre-computed S-Box masking, making masking countermeasures an unlikely choice for TinyML deployments ~\cite{sbox}. Other research has shown that masking, in the worst case, can cost more than seven times more compute cycles and eleven times more RAM utilization ~\cite{power_analysis_dis}. It has also been shown that constant time activation functions (a subclass of hiding) require significantly more computational overhead, with the multiplication operations taking twice as much time~\cite{phys_SCA}. More research should be done to benchmark how these defenses affect the power consumption and latency of TinyML devices and to create more efficient alternatives.
    \item \textbf{Voltage Regulators: (SCA)} Voltage regulators are an expanding area of side-channel defense research that offers lower cost options that may be feasible on resource-constrained TinyML devices. The authors in~\cite{kar_invited_2017} advocate using integrated voltage regulators (IVRs) in IoT and edge devices that can isolate the local supply nodes of the digital logic from the input nodes making it more difficult for attackers to obtain useful power traces. This technique requires specialized packing but can be low-power and chip-area efficient. Chen et al~\cite{chen_volt} advocate further confusing power and electromagnetic attacks by dynamically adjusting the voltage. Their results show that their implementation of Island-based Random Dynamic Voltage Scaling prevented attackers who had two hundred thousand encryption traces from recovering secrets. Although the results from these studies are promising, many voltage regulation techniques have yet to be adopted by commodity microcontrollers, making these countermeasures out of reach for many developers.
    \item \textbf{Memory Encryption (Memory Extraction):} Memory encryption and the use of TEEs such as Intel SGX provide robust defenses against attackers physically accessing device memory, however, may not be feasible for TinyML class devices. Unlike Intel SGX, Arm TrustZone (offered on many embedded processors) does not encrypt enclave data, making it susceptible to attackers tapping the bus or memory to recover plain text data~\cite{SGX}. In addition, TrustZone is not offered on TinyML class MCUs (ARM Cortex M0, M0+), only higher-end Arm Processors~\cite{ARM_TRUSTZONE}. Outside of TEEs, memory encryption is also costly and most likely outside the scope of TinyML due to the computational and area overhead as well as the power consumption associated with constantly encrypting and decrypting models and data points. With the proliferation of these technologies and their physical accessibility, more research should be done to produce defenses that offer similar protections to those mentioned above at a cheaper cost.
    \item \textbf{Secure Boot (Memory Flashing Attacks):} Secure boot is an essential feature to establish a hardware root of trust and prevent malicious code injection or application spoofing. Luckily, this feature is readily available on commodity MCUs and does not require substantial overhead. 
    \item \textbf{Redundant Circuitry / Computation (FIAs):} These defenses offer a start at defending against FIAs, but are not a cure-all due to the diversity of this class of attacks. Even so, these countermeasures may not be a viable option for deeply embedded MCUs. Redundant circuitry comes at a significant area cost that is not palatable for TinyML class devices~\cite{fia_counter}. The same can be said for redundant computation as the computational overhead of ML inference would drastically increase with redundant operations, increasing latency and power consumption unless heuristics that reduce robustness are used~\cite{fia_counter}. 
    \item \textbf{RDP, PCROP, WRP (Memory Extraction, Memory Flashing):} These countermeasures are all viable options for TinyML class devices as they are implemented on all STM32 devices. However, they should not be used as a holistic security solution because their functionalities vary by device. RDP provides read protection for flash memory in all cases, but not for SRAM on more lightweight devices such as Arm Cortex-M0+ MCUs~\cite{STM32MCUsSecurity2024}. This means that attackers may still be able to dump data from SRAM on lightweight STM32 architectures. PCROP also has varying functionality depending on the MCU used. In some cases, for Cortex-M0+ MCUs, there are only options to protect up to 512 bytes of memory, leaving large portions of code vulnerable to being written to or read ~\cite{STM32MCUsSecurity2024}. The same is the case for WRP where on Cortex-M0+ MCUs, you can only protect up to 4KB of flash and no SRAM ~\cite{STM32MCUsSecurity2024}. This is all to say that many deeply embedded MCUs do not have adequate built-in memory protections. Further research is required in this area to develop more robust and efficient memory protection mechanisms for deeply embedded MCUs. In the meantime, developers should thoroughly review security documentation before settling on an MCU to use.
    \item \textbf{ECC and PVD (FIAs):} Again, both of these countermeasures are available on all commodity STM32 MCUs but should not be a holistic solution. The ECC mechanism is defined as a ``safety feature" and a ``complementary protection" against FIAs in the STM documentation meaning it likely does not offer very high robustness to these attacks~\cite{STM32MCUsSecurity2024}. PVD on the other hand does offer a reasonable amount of robustness by allowing a configuration that detects when voltage drops below a threshold value, raises an interrupt, and cleans memory~\cite{STM32MCUsSecurity2024}. However, this countermeasure does not cover fault delivery mechanisms other than voltage glitching. These being the only available options for FIA defenses on deeply embedded devices is due in large part to the high overhead of more robust methods. More research should be done to develop resource efficient methods to secure deeply embedded commodity MCUs.
\end{enumerate}

\section{TinyML Software Security Challenges} 
\label{sec:software-challenges}
The niche TinyML fits leaves very little opportunity for attackers to exploit these devices remotely. That is because, again, the primary purpose of these devices is to produce real-time inference on the deep edge and relay outcomes. As a result, there is no reason for these devices to have services that users can interact with or exploit. Because this paper is focused on TinyML devices specifically, we will not address attack vectors present in the IoT network infrastructure that TinyML uses. Instead, we will focus on attacks that can be mitigated by controls that can be implemented on a TinyML device. With this scoping, the most probable ways an attacker could exploit these devices remotely is through vulnerable communication or model update mechanisms.

\subsection{TinyML Communication Channel Attacks}
\textbf{Overview:} A key part of the Edge device workflow is transmitting data or inference outcomes through an IoT network infrastructure to be consolidated on back-end servers. There are several confidentiality and integrity concerns associated with this process. On standard, more robust systems, attacks that take advantage of these communication channels can be mitigated trivially by using defenses such as TLS. However, TinyML class devices must use less robust countermeasures such as those built into low-power protocols like LoRaWAN.

\textbf{Attack Description:} In the context of confidentiality, an attacker can eavesdrop on plain-text packets to steal the artifacts being transmitted by these devices. Integrity concerns include Man-in-the-middle (MITM) type attacks, particularly spoofing attacks where an attacker can modify the packet in transit or impersonate the device transmitting artifacts~\cite{mitm}. In the context of TinyML, these attacks can exploit insecure devices to steal inference outcomes, modify inference outcomes, or flood the receiving server with arbitrary inference results. Our inputs for eavesdropping attacks are adjacent network attack vector (attacker must be collocated on the same physical or logical network, e.g., cannot eavesdrop across router hops), low attack complexity, no privileges required, no user interaction, unchanged scope, and high confidentiality risk. Similarly, MITM attacks are an adjacent network attack vector, have high attack complexity, no privileges are required, no user interaction, unchanged scope, and high confidentiality and integrity risk. As a result, eavesdropping attacks receive a medium CVSS score of 6.5 and MiTM attacks receive a high CVSS score of 7.5.

\textbf{Traditional Countermeasures}: These are well-studied vulnerabilities that can be mitigated trivially in the context of EdgeML and IoT because the computational cost of cryptography is not as prohibitive on more robust devices. That is, the most common way to mitigate these kinds of attacks is through using Transport Layer Security (TLS) which makes use of both symmetric and asymmetric cryptography. By encrypting the message contents, TLS prevents an attacker from eavesdropping or modifying packet contents in transit~\cite{tls}. TLS clients and servers can also optionally make use of X.509 certificates to allow mutual authentication to prevent spoofing attacks~\cite{tls}. These countermeasures are always options on robust devices such as personal computers and smartphones, and are usually readily available on more robust IoT devices that make use of services such as AWS IoT Core~\cite{AWS_docs}. However, more resource-constrained edge devices, like those of TinyML, are usually unable to make use of the traditional TCP/IP networking stack both due to computational overhead and the use cases of these devices~\cite{auth_survey}. As a result, these devices often must use IoT-specific protocols such as Bluetooth Low Energy (BLE), ZigBee, and LoRaWAN~\cite{iot_protos}. 

\textbf{Considerations for TinyML Countermeasures}: Due to the constraint that these protocols must run on resource-constrained devices, the security options these protocols provide are often less robust than TLS. BLE is a widely adopted wireless communication mechanism used on resource-constrained edge devices due to reduced power consumption~\cite{iot_protos}. BLE provides optional confidentiality, integrity, and authentication protections through the use of Elliptic Curve Diffie-Hellman (ECDH) and AES. However, over the years through many revisions, BLE has remained vulnerable to MITM and eavesdropping attacks due to weaknesses in its pairing methods~\cite{iot_protos, ble_vulns}. Zigbee is another popular protocol in the edge computing space that is more robust than BLE, but still has its faults. Zigbee makes use of AES to encrypt and authenticate communication between devices. However, only the payload is encrypted, not the device address, allowing Denial of Service attacks that aim to deplete the device battery by flooding the device with network packets~\cite{iot_protos}. LoRaWAN is another popular networking protocol that has been widely adopted for TinyML and other Edge use cases because it offers long-range communication with low-power consumption, unlike protocols such as Zigbee and BLE~\cite{lora_sec_short}. Devices using LoRaWAN are deployed with a pre-configured ``AppKey" used to derive an``AppSKey" and ``NwkSKey" used for payload encryption and generating message integrity codes using AES, respectively~\cite{lora_sec}. Many of the well-known vulnerabilities for LoRaWAN such as replay attacks, eavesdropping, ack spoofing, and denial of service have been patched as of version 1.1~\cite{lora_backwards}. However, LoRaWAN 1.1 offers a backward compatibility mode that allows devices to operate with earlier versions of the protocol, which may still be susceptible to some of these attacks~\cite{lora_backwards}.

\begin{openquestionbox}
\begin{itemize}
\it 
\item What are the specific memory and power costs associated with the security features provided by protocols such as LoRaWAN, and to what extent are these costs sustainable within the constraints of TinyML?
\item How does memory sharing between ML models and active software countermeasures impact system performance in TinyML devices?
\item To what degree does the overhead introduced by these countermeasures affect inference latency?
\end{itemize}
\end{openquestionbox}

\subsection{Model Update Attacks}
\label{subsec:update}
\textbf{Overview and Background:} Model updates are crucial for any ML system. They allow a device to adapt to its environment and keep pace with advancements in TinyML and ML as a whole. However, if implemented incorrectly, model update exploits can have catastrophic consequences. The severe resource constraints inherent in TinyML deployments often mean that these devices do not run fully-fledged operating systems. As a result, the application and system firmware consists of a single firmware image~\cite{TinyML_updates}. Thus it is difficult for these devices to be updated in pieces, rather, they must be re-flashed with an entirely new image~\cite{TinyML_updates}. However, in more recent work, researchers have refactored the popular IoT operating system RIOT to be usable in TinyML applications and have labeled it as RIOT-ML~\cite{riot_tinyML}. One of the major contributions of this work is providing a mechanism for developers to update their TinyML devices with more granularity. For example, developers may only update the model or even a single layer on their device~\cite{riot_tinyML}.
These updates are usually done using an over-the-air (OTA) medium regardless of the update system or library used as it is laborious to do this through physical or proximity access~\cite{TinyML_updates}. 

\textbf{Attack Description} The Minerva update system showcased in~\cite{TinyML_updates} is simple from a security perspective like many other OTA update systems used for TinyML devices. It relies solely on checksum integrity checks to ensure no bits were dropped or flipped in transit. That is, there are no mechanisms to protect against adversarial data corruption or to authenticate the person sending the update to the device. The lack of these protections creates several opportunities for attackers to exploit OTA update-enabled TinyML devices. First, since the firmware update is sent over the wire in clear text, a passive attacker could sniff the traffic and steal the intellectual property contained in the device firmware ~\cite{mitm}. Second, and more concerning, since there are no authentication checks, an attacker who knows the semantics of the update system could spoof the updates and send their own update to make the device behave arbitrarily ~\cite{mitm}. For example, to breach the device's confidentiality, an attacker could simply send a firmware update that routes all data collected by the device to some attacker controlled server. An attacker could alter the integrity of the device by replacing the intended ML model with another model that behaves in a way that allows an attacker to attain their goals. Finally, the attacker could impact the availability of the device by sending a firmware update that bricks the model. The CVSS inputs for this example attack are an adjacent network attack vector, high complexity, no privileges required, no user interaction, unchanged scope, and high confidentiality, integrity, and availability risks, resulting in a high CVSS score of 7.5.

\textbf{Countermeasure Overview:} The three security primitives to keep in mind to defend against these kinds of attacks are confidentiality, integrity, and authenticity. For confidentiality protection we want to ensure that any information coming into or going out of the device cannot be read in plain text by a passive attacker; for authenticity and integrity, we want to ensure that the firmware update is coming from a trusted source and cannot be modified in transit. The Internet Engineering Task Force (IETF) standard for OTA model updates is the SUIT protocol described in RFC 9019~\cite{rfc9019}. This protocol provides integrity and authenticity checks by default and optional confidentiality protections. The SUIT authentication mechanism uses asymmetric cryptography to generate trusted cryptographic signatures server side and to verify them device side. The integrity mechanism consists of verifying manifest checksums to ensure no bits were dropped or flipped either in transit or maliciously. More recent work by the IETF has provided a specification for encrypting the payloads of SUIT manifests using ephemeral-static Diffie-Hellman~\cite{rfc9052} or AES Key Wrap~\cite{rfc3394}, which provides confidentiality protection for firmware updates~\cite{suit_encryption}. SUIT and the associated semantics have been implemented in RIOT OS and ported over to RIOT-ML, making it available for use in TinyML deployments~\cite{riot_tinyML}. In totality, the network stack, SUIT implementation, and required cryptographic libraries consume about 55 KB of storage~\cite{riot_tinyML}, making it a viable option to ensure secure updates in TinyML deployments. However, it should be noted that the introduction of a larger code base through the usage of an embedded OS could introduce more vulnerabilities to the TinyML device. In addition, again, the use of ephemeral-static Diffie-Hellman or AES Key wrap may introduce unpalatable computational overhead and latency. More information about the semantics of the SUIT implementation in RIOT-ML and associated benchmarks can be read about in~\cite{riot_tinyML}.

\begin{openquestionbox}    
\begin{itemize}
\it 
\item How can we effectively assess and validate that the additional functionalities provided by RIOT-ML's model update framework do not inadvertently expand the attack surface for TinyML devices?
\item What are the specific interactions between RIOT-ML's controls and functionalities and the ML model in TinyML implementations?
\item To what extent do these interactions impact inference latency and power consumption?
\end{itemize}
\end{openquestionbox}

\subsection{Software Countermeasure Viability for TinyML}

Through this section, we have covered several countermeasures that can be used to secure network-facing TinyML devices. Many of these defenses hinge on the use of cryptography which is known to be computationally expensive. Here, we assess the feasibility of these controls in the context of TinyML. For our analysis, we use code and library sizes to speculate about the implications these controls may have for TinyML class devices. Recall that these devices must also continuously run firmware, a networking stack, collect data from a sensor, and run ML inference in addition to the proposed countermeasures. For more concreteness, in a study by Banbury et al.~\cite{micronets}, the authors recorded the latency, memory, and storage metrics associated with running various models for different TinyML tasks. Of note, they found that running DS-CNN for keyword spotting (KWS) requires 216KB of SRAM and 687 KB of flash, leaving only 102KB of SRAM free and 312 KB of flash on a device with 320 KB of SRAM and 1MB of flash~\cite{micronets}. As a result, there will be fairly limited memory and storage resources remaining to run miscellaneous applications, networking stacks, and countermeasures.

\begin{enumerate}[noitemsep]
    \item \textbf{TLS (Eavesdropping, MITM):} TLS is the foundation of secure communication over the modern internet. However, it may not be able to run trivially on TinyML class devices. TLS depends on asymmetric encryption for authentication, Diffie-Hellman Key Exchanges, and symmetric session keys~\cite{tls}, all of which require non-trivial amounts of memory overhead. Perhaps the best off-the-shelf TLS implementation for Edge devices is Mbed TLS, which offers access to X.509 certificate usage and TLS/DTLS~\cite{mbed_tls}. According to their documentation, Mbed TLS can have a memory footprint ranging from 45KB to 300KB depending on the features your implementation requires~\cite{mbed_tls}. On top of the code footprint, Mbed TLS has a default buffer size of 16KB. In order to make use of Mbed TLS, you will also need a networking stack such as LWIP which requires tens of KBs of RAM and 40KB of ROM~\cite{lwip}. More research is required to benchmark the RAM and storage requirements of this configuration on TinyML class devices, however, given the overhead of the ML model, it may be difficult, if even possible, to implement Mbed TLS on TinyML devices. In addition, if implemented, Mbed TLS may lead to unpalatable latency and power consumption. 
    \item \textbf{AES (Eavesdropping, MITM):} AES is the encryption algorithm used by all the resource-constrained networking protocols discussed previously. As such, it is usually feasible to run on TinyML class devices. AES libraries that are optimized to be small such as the implementation shown in~\cite{tiny_aes} have a code footprint of less than 1KB. The study done in~\cite{aes_wsn} shows that the encryption and decryption operations done on TinyML class devices take time on the order of microseconds, meaning the use of AES is unlikely to introduce unwanted latency. However, more research should be done to benchmark and observe the effects of ML inference and AES competing for limited computational resources.
    \item \textbf{ECDH (Eavesdropping, MITM):} ECDH is another key security control used in resource-constrained networking protocols to derive shared secrets. ECDH is a reasonable control to implement on TinyML class devices due to a small code size of just over 1KB~\cite{tiny_ecdh} and an execution time on the order of milliseconds~\cite{ecdh_bench}. Again, more research should be done to observe how ECDH, AES, ML inference, and other competing processes interact with the limited pool of computational resources.
    \item \textbf{RIOT-ML (Model Update Attacks):} As a tool suite designed for TinyML devices, RIOT-ML presents a start at creating more homogeneous, adaptable TinyML devices~\cite{riot_tinyML}. However, as a new library that has not been extensively vetted by the open-source community, developers should be aware that there may be unforeseen security challenges created RIOT-ML due to a larger set of capabilities that may be exploited.
\end{enumerate}

\begin{figure}[htbp]
    \centering
    \includegraphics[width=0.5\textwidth]{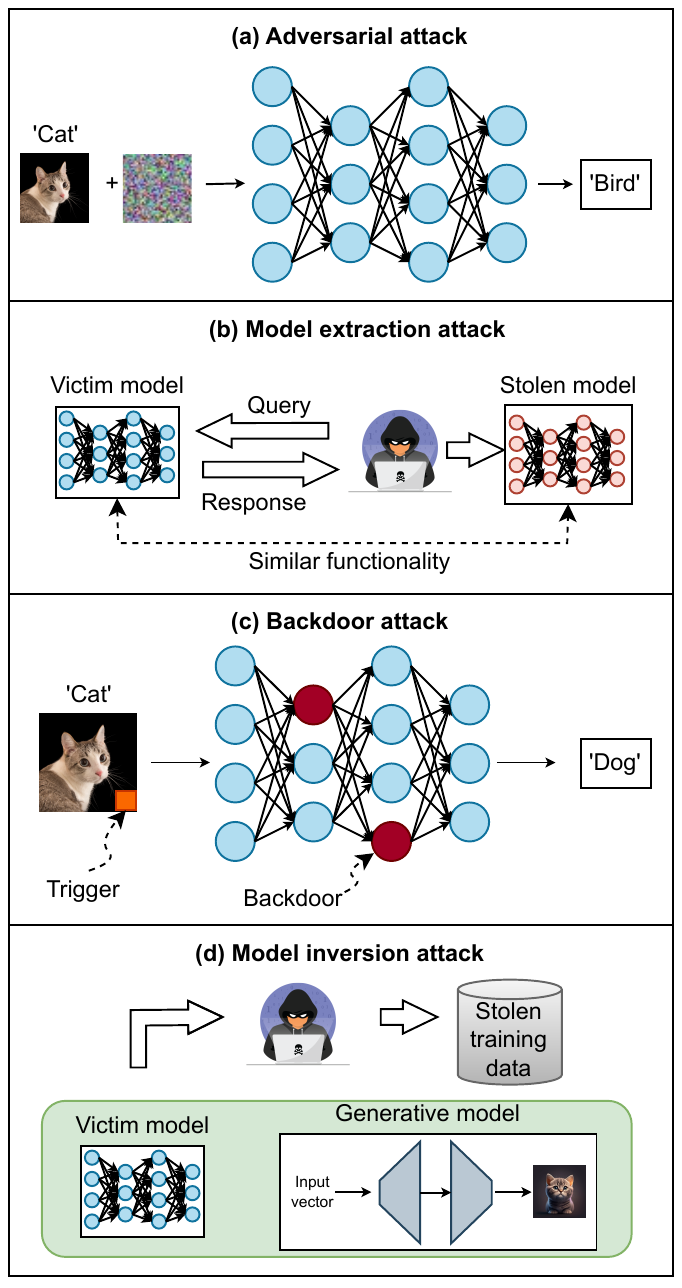}
    \caption{TinyML Machine Learning Security Challenges Overview. The primary classes of attacks on machine models and high level descriptions of how they are carried out are depicted.}
    \label{fig:ml_challenges}
\end{figure}

\section{TinyML Model Security Challenges}
\label{sec:ml-challenges}
Outside of traditional software exploits, there are several ways that the TinyML model could be exploited through traditional attacks on larger ML models such as adversarial example attacks, model extraction attacks, model inversion attacks, and backdoor attacks. Here we discuss these attacks, their countermeasures, and how each of these change due to TinyML implementation semantics as compared to the EdgeML setting. A high-level overview of the attacks we cover in this section can be found in Figure \ref{fig:ml_challenges}. We summarize the attacker's knowledge required for various types of attacks—adversarial, model extraction, backdoor, and model inversion—in Table~\ref{tab:algo_attack_threat_model}. While the specific knowledge needed varies by attack type and scenario, it generally includes access to inference inputs, model architecture and parameters, model outputs, training data, or subsets thereof. For TinyML systems, these pieces of knowledge can potentially be acquired by malicious attackers. Attackers might gain access to and poison inference inputs through hardware attacks. The model architecture, parameters, and training data could be publicly released or left unprotected. Moreover, the training process could be exposed during execution on third-party platforms.

\begin{table}[t]
    \centering
    \resizebox{\columnwidth}{!}{
    \begin{tabular}{|c|c|c|c|c|c|c|}
        \hline
        \multirow{2}{*}{Threats} &  \multirow{2}{*}{\makecell[c]{Attack \\ setup}} & \multicolumn{5}{c|}{Attacker's Knowledge} \\ \cline{3-7}
        
        & & \makecell[c]{Inf. \\ input} & \makecell[c]{Model arch. \\ \& param.} & \makecell[c]{Model \\ output} & \makecell[c]{Train. \\ data} & \makecell[c]{Train. \\ process} \\ \hline
        
         \multirow{2}{*}{Adversarial} & w  & \ding{51} & \ding{51} & \ding{51}  &  & \\ \cline{2-7}
         
          & b  & \ding{51} &  & \ding{51}  &  & \\ \hline
          
         \multirow{2}{*}{Model Extraction} & w  & \ding{51} & \ding{51} & \ding{51}  &  & \\ \cline{2-7}
         
          & b  & \ding{51} &  & \ding{51}  &  & \\ \hline
        
        Backdoor &  & &  &  & \ding{51}   & \ding{51}  \\ \hline
        
        \multirow{2}{*}{\makecell[c]{Model \\ inversion}} & w  &  & \ding{51}  & \ding{51}   &  &  \\ \cline{2-7}
          & b  &  &  & \ding{51}  &  &  \\
        \hline

    \end{tabular}}
\caption{Knowledge required for different threats. 'w' indicates white-box setup and 'b' indicates black-box setup.}
\label{tab:algo_attack_threat_model}
\end{table}

\subsection{Adversarial Example Attacks}
\textbf{Overview:} A well known type of attack on ML models are \textit{adversarial example attacks}. This class of attacks aims to thwart model integrity by making them misclassify perturbed inputs. In the TinyML setting, this could have devastating consequences in health care and autonomous vehicle deployments. We find that the most realistic way to combat these attacks is with training-based defenses, while pre-processing based defenses often require more computational overhead than what is palatable for TinyML devices. 

\textbf{Attack Description:} In adversarial example attacks, a malicious actor adds small perturbations to a model input with the goal of having the model produce an inaccurate output~\cite{benchmarking_adv}. For instance, Eykholt et al. demonstrated that perturbations applied to physical road signs could lead to a convolutional neural network (CNN) misclassifying the altered images with up to 100\% confidence~\cite{physical_adv}. Adversarial examples can either be targeted or untargeted, meaning that an attacker can attempt to forge the input such that it will be classified into a target class that is not the intended class, or, into any class that is not the intended class~\cite{benchmarking_adv}. There are several attack methods that can be used to achieve these ends such as DeepFool~\cite{deepfool}, AutoAttack~\cite{autoattack}, Zeroth Order Optimization (ZOO)~\cite{zoo}, and Geometric Decision-Based Attack (GeoDA)~\cite{geoda}. These attacks can be particularly devastating for TinyML use cases in fields such as healthcare or autonomous vehicles. In these scenarios, the integrity of the ML model's performance is absolutely paramount as erratic behavior can mean life or death for users. An adversarial attack in health care may lead to a misdiagnosis while an adversarial attack on an autonomous vehicle may cause it to miss a road sign, causing an accident. Although these attacks can have devastating consequences, in practice, they are often very difficult to execute in physical environments. This is because of varying model architectures, differing sensing distances and angles, varying backgrounds, and sensor imperfections~\cite{physical_adv}.

\textbf{Considerations for TinyML:} These attacks are particularly interesting for TinyML devices for the following reasons.

\begin{enumerate}[noitemsep]
    \item \textbf{Size of the Model:} TinyML models fundamentally have fewer layers and weights than larger language models due to resource constraints. These model footprints are further reduced by methods such as quantization. The tiny nature of these models and the resource constraints associated with these devices change how we should approach defending them. 
    \item \textbf{Environment:} The deployment of TinyML devices into the physical world introduces unique challenges and vulnerabilities. Unlike many ML models that are primarily accessed through software APIs or other prompting mechanisms, TinyML devices interact directly with their physical environment through sensing.
\end{enumerate}

TFLM, now a part of Google AI Edge, is the most used framework for deploying TinyML models~\cite{TFLM}. TFLM provides developers with a toolkit to train, optimize, and deploy small ML models on resource-constrained microcontrollers. The current optimization options offered by TFLM are quantization, pruning, and clustering~\cite{TFLM_docs}. Full integer quantization is a necessity in TinyML use cases as it makes the model up to four times smaller and at least three times faster, while an optimization option like pruning allows more efficient model compression but does not reduce its memory footprint~\cite{TFLM_docs}. Some previous research has been done on the effects of quantization on a model's adversarial robustness in~\cite{adv_1},~\cite{adv_2},~\cite{adv_3},~\cite{Costa2024DavidEdge}. In research done by Bernhard et al., the authors found that models quantized to 1,2,3, and 4-bit precision performed very similarly to the full-precision 32-bit models when tested on adversarial examples, although quantization in and of itself is a weak defense against adversarial examples~\cite{adv_1}. These findings are consistent with the work done in~\cite{adv_2} and~\cite{adv_3}, where the authors found that models that were optimized with post-training quantization (PTQ) were within 5\% in performance with the non-quantized models and that the quantized neural networks (QNNs) tested were robust to 98.07-99.69\% of the adversarial examples provided to the corresponding full-precision model, respectively. Perhaps the most important paper written on this topic was done by Costa and Pinto as it looked at the effects of quantization on adversarial robustness in the context of TinyML, using TFLM~\cite{Costa2024DavidEdge}. These authors found that in general, 8-bit quantized models are more robust to 4 different adversarial attacks (ZOO, Square Attack, Boundary Attack, and GeoDA) than their 16-bit and full-precision counterparts~\cite{Costa2024DavidEdge}. Given the findings in these papers, we can be confident that without any other adversarial defenses QNNs will be about as robust to adversarial attacks as their full-precision counterparts. The CVSS inputs for this attack are physical attack vector, high attack complexity, no privileges required, no user interaction, unchanged scope, and high integrity risk, giving it a medium CVSS score of 4.2.

\textbf{Countermeasure Overview:} Costa and Pinto also looked at the effectiveness of applying common defenses for adversarial examples used on larger models to QNNs. They evaluated four train-based defenses, Defensive Distillation~\cite{def_distill}, PGD Adversarial Training~\cite{pgd_train}, Ensemble Adversarial Training~\cite{ensem_training}, Sinkhorn Adversarial Training~\cite{sat_adv}, and two pre-processing based defenses, Feature Squeezing~\cite{feat_squeeze} and Pixel Defend~\cite{pix_def}~\cite{Costa2024DavidEdge}. In their evaluation, they found that defenses that are viable for full-precision models are also viable in their less precise QNN implementation~\cite{Costa2024DavidEdge}. However, the significant overhead of the pre-processing based defenses mean they may not be viable for many TinyML deployments~\cite{Costa2024DavidEdge}. Finally, they found that PGD and Sinkhorn adversarial training present the best options for creating robust QNNs~\cite{Costa2024DavidEdge}. 

While these proposed model-based defenses provide a foundation for enhancing the robustness of TinyML devices through model-level protections, TinyML deployment environments expose these devices to different attack vectors that are less prevalent in larger, cloud-based models. TinyML models rely on the data collected by an on-board sensor while cloud models will take input through some API or prompting mechanism. As such, the integrity of the ML model relies heavily on the integrity of the data collected and passed to it by the sensor. In ~\cite{asia_talk_paper}, the authors provided a proof-of-concept adversarial example attack on a time-series classification TinyML device. This proof-of-concept attack added small perturbations to the data provided to the model resulting in 95.76\% attack efficacy at the highest attack strength. While the adversarial example defenses enumerated in Costa and Pinto would provide some robustness to this kind of attack, a truly robust TinyML system should provide other protections to ensure data integrity and deny persistent provisioning of adversarial examples. These protections include those listed previously in sections ~\ref{subsubsec:flash} and ~\ref{subsec:update} that protect the integrity of the pre-processing code and data in memory. Implementing and testing these adversarial example defenses in each unique TinyML implementation is essential to maintaining device integrity by providing resistance to adversarial examples.

\begin{openquestionbox}
\begin{itemize}
\it
\item To what extent can pre-processing based defenses be developed and optimized specifically for resource-constrained QNNs to enhance their resilience against adversarial example attacks in TinyML?
\item What are the quantitative impacts of implementing pre-processing based defenses, suitable for TinyML applications, on inference latency and power consumption?
\item How do these impacts vary across different TinyML device configurations?
\end{itemize}
\end{openquestionbox}
\subsection{Model Extraction Attacks}

\textbf{Overview:} Another well known attack on ML models is a \textit{model extraction attack}. In model extraction attacks, the attacker takes advantage of query access to create input/output pairs to make a functionally equivalent model ~\cite{extraction}. These attacks are particularly interesting because they can be used as an initial exploit to make adversarial example and model inversion attacks easier. We hypothesize that these attacks may be easier to carry out on TinyML models due to their smaller size, limited pool of operations, and the widespread usage of pre-trained models. Additionally, traditional countermeasures for these attacks, such as query throttling and monitoring may be more difficult to apply in the TinyML setting due to limited computational resources.

\textbf{Attack Description:} In a model extraction attack, the attacker is able to give the target model inputs of their choosing and observe the model outputs. There are many methods to conduct this kind of attack but perhaps the most well-researched method and least restrictive is using a learning-based approach~\cite{extraction}. Using a learning-based approach, an adversary may either use a synthetic unlabeled dataset such as in~\cite{practical_black_box} or a publicly available dataset like ImageNet~\cite{extraction}. These datasets are used to query the model and allow the attacker to collect the outputs for use as labels to train a functional equivalent. The difficulty of these attacks depends on a few factors including the size of the target model, the adversary's domain knowledge, deployment knowledge, and model access~\cite{extraction}. Domain knowledge refers to what the attacker knows about the task the model was designed for, deployment knowledge is what the attacker knows about the target model itself, and model access is what metrics or artifacts the attacker receives from the model output~\cite{extraction}. In~\cite{asia_talk_paper}, the authors demonstrated a proof-of-concept model extraction attack on two different TinyML devices, one being a computer vision use case and the other being gesture recognition. Using a learning-based approach, they were able to create extracted models with 90\% and 69.23\% accuracy, respectively~\cite{asia_talk_paper}. On its face this attack poses a threat to device confidentiality as it allows an attacker to steal the intellectual property in the model itself. However, this attack can also be used to guide adversarial example attacks by allowing attackers to craft transferable adversarial examples~\cite{practical_black_box}, or to make model inversion attacks easier given information about the target model~\cite{extraction_apis}, adding potential device integrity and privacy considerations.

\textbf{Considerations for TinyML:} While there has been extensive research on model extraction attacks targeting machine-laerning-as-a-service (MLaaS), there has not been any research, aside from~\cite{asia_talk_paper}, looking at these attacks applied to TinyML specifically. However, there are a few features of TinyML that speculatively make model extraction attacks easier on these devices.

\begin{enumerate}[noitemsep]
    \item \textbf{Model Size and Limited Supported Operations:} While larger models that are used for MLaaS have a memory footprint on the order of gigabytes, TinyML model footprints are only megabytes. In addition, the popular TinyML library TFLM only supports 130 operations while the full TensorFlow library supports over 1,400, further reducing the number of possible model configurations~\cite{TFLM}. This is all to say that the search space for these model extraction attacks may be significantly smaller than those for larger models, making these attacks easier.
    \item \textbf{Use of Pre-trained Models:} The use of pre-trained models has become common practice in the machine learning industry to reduce costs and take advantage of state-of-the-art models. This is also the case in TinyML, where developers want to maximize accuracy and minimize memory footprint. Because it is so difficult to come up with novel model architectures that fit these two constraints, many pre-trained models such as MobileNetV1, SqueezeNet, and LeNet-5 are commonly used in TinyML deployments~\cite{datasheets},~\cite{alajlan2022tinyml}. This fact increases an attacker's deployment knowledge, given they have reasonable domain knowledge, making model extraction attacks even easier for TinyML devices. 
\end{enumerate}

These two heuristics speculatively make model extraction attacks easier on TinyML devices, however, there are also a few other heuristics that make model extraction attacks harder on these devices. Firstly, these attacks require physical access to the device and a way to manipulate the data being bussed to the MCU from the sensor. This makes it more difficult to query TinyML models than it is to query MLaaS models through some prediction API. Secondly, the outputs of a TinyML device ought not to be in plain text until it reaches either a back-end server or some application processor, while MLaaS models must output plain text to be usable. The CVSS inputs for this attack are physical attack vector, high attack complexity, no privileges required, no user interaction, unchanged scope, and high risk for confidentiality, giving this attack a medium score of 4.2 on the CVSS scale.

\textbf{Countermeasure Overview:} These nuances dictate the way we should go about defending these devices. In traditional settings, model extraction defenses usually aim to either limit the amount of information given by the model response, limit API accesses or detect malicious query patterns~\cite{extraction},~\cite{extraction_apis}. It is also best practice to limit the information given by the model output for TinyML use cases unless other artifacts such as confidence scores and logits are necessary. However, it may be more difficult to detect malicious query patterns or limit inference access on TinyML devices. The overhead required to monitor query patterns may not be plausible due to resource constraints while how often the device makes an inference may deteriorate the device's utility. One unique way to defend these devices from model extraction attacks is by encrypting the device output. Unlike traditional MLaaS systems, TinyML devices relay their inference outcomes to some back-end server or application processor to be aggregated and computed over. Thus, in order to complete the query-output pair to train the functional equivalent, the attacker would also need to exploit the system the TinyML device delivers its inference outcomes to. Future research is required to test the viability of existing controls, but for now, encrypting device outputs should make model extraction attacks harder to execute on TinyML devices.

\begin{openquestionbox}
\begin{itemize}
\it
\item How does the search space of model extraction attacks on TinyML models compare to that of larger models? 
\item What are the implications of these differences on the cost and feasibility of such attacks in  TinyML?
\item To what extent can efficient query monitoring protections be implemented on TinyML devices without significantly impacting their inference procedures? 
\item What are the trade-offs between protection efficacy and performance impact?
\end{itemize}
\end{openquestionbox}

\subsection{Backdoor Attacks}
\textbf{Overview:} Backdoor attack is a type of adversarial attack on machine learning models, where an attacker embeds hidden triggers or malicious patterns in the training data. The model behaves normally on regular inputs, but when the trigger is present, it produces a specific, attacker-chosen output.

\textbf{Attack Description:} Backdoor attacks have been proven effective on TinyML networks such as LeNet-5. BadNets~\cite{gu2019badnets} demonstrates that inserting a colored square into a subset of the training data and conducting malicious training can effectively embed a backdoor task into a model. 
The Trojaning attack~\cite{liu2018trojaning} avoids the need for the original training data by reverse-engineering the training data through gradient descent and denoising, followed by malicious retraining with the generated data.
The Blend backdoor attack~\cite{chen2017targeted} demonstrates that a backdoor can be injected using only about 50 poisoning samples while achieving an attack success rate of above 90\%.
The Clean-label backdoor attack~\cite{turner2018clean} leverages GAN-based interpolation and adversarial perturbations to poison inputs so that they appear consistent with their labels and benign upon human inspection. A backdoor attack without label poisoning is proposed in~\cite{barni2019new} to increase the stealthiness of the attack, as samples whose content does not agree with the label in the manipulated training set can be identified by visual inspection or a pre-classification step. The Dynamic backdoor attack~\cite{nguyen2020input} develops an input-aware trigger generator, enabling triggers to vary from input to input.
WaNet~\cite{nguyen2021wanet} uses a small and smooth warping field to generate backdoor images, making the modifications less noticeable than adding patch perturbations such as noise, strips, or reflectance.
The Clean-label poisoning attack~\cite{shafahi2018poison} crafts poisoned images by colliding them with a target image in the feature space, making it difficult for a network to discern between the two.
The Deep Feature Space Trojan Attack~\cite{cheng2021deep} proposes a controlled detoxification trojaning method to ensure that the injected backdoor depends on deep features rather than shallow ones. Deep features, extracted from deeper layers of a neural network, capture more abstract and complex representations of the data, while shallow features, derived from earlier layers, represent basic and simple attributes. This method also indicates that increased resource consumption during the trojaning process leads to more difficult-to-detect trojaned models.
LIRA~\cite{doan2021lira} unifies the process of generating trigger patterns and poisoning the model under a single non-convex constrained optimization framework. The resulting triggers are stealthy and dynamic, successfully poisoning a classifier while maintaining unchanged performance on clean data and achieving high attack success rates.
Adaptive-Blend~\cite{qi2022circumventing} studied the latent separation of backdoor triggers and designed an adaptive backdoor attack that can actively suppress the latent separation and thus circumvent latent separation-based defenses~\cite{hayase2021spectre}.

In addition to traditional backdoors, recent research has unveiled a novel attack paradigm known as conditioned backdoors, which remain dormant within a model until activated by specific post-training processes like pruning, model quantization, or fine-tuning for downstream tasks. Given that pruning, quantization, and fine-tuning are essential for adapting neural network models to the resource limitations of TinyML devices, conditioned backdoor attacks pose a significant new security challenge in the TinyML landscape.
~\cite{tian2022stealthy} first designed conditioned backdoor attacks for both model quantization and model pruning. This work assumes the attacker knows the specific quantization and pruning methods which is a strong assumption. 
~\cite{pan2021understanding} pointed out that it is risky to deploy third-party QNNs on edge devices without defenses by showing that the existing backdoor defensive techniques designed for detecting the backdoors in the full-precision neural network are ineffective for QNNs. 
~\cite{hong2021qu} first investigated the transferability of a quantization-conditioned backdoor attack where the victim uses a different quantization scheme than the attacker considered. The effects these attacks have on the model's integrity earn them a medium CVSS score of 4.2.

\textbf{Countermeasure Overview:} The defenses of backdoor attacks can be categorized into two categories: (1) disabling the trigger, and (2) pruning the backdoor neurons in the network.

It has been revealed that modifications in the location or appearance of the trigger markedly impair the performance of backdoor attacks~\cite{li2020rethinking}. This insight has been leveraged to formulate a defense against backdoor attacks that utilize static triggers, involving slight alterations in the inputs' location and appearance to deactivate the backdoors~\cite{li2020rethinking}.
~\cite{tran2018spectral, chen2018detecting, liu2022complex} use a similar strategy by using statistical deviation in the feature space.~\cite{wang2019neural, hu2021trigger} tried to recover the trigger pattern and defend against the backdoor attack by detecting and blocking the poisonous inputs.
Additionally, it is observed that backdoored networks contain both clean and backdoor neurons. The backdoor neurons tend to be dormant when the inputs are benign samples and be activated only by trigger patterns~\cite{liu2018fine}. Based on this, a line of work was proposed to prune the backdoor neurons in a backdoored network. Fine-pruning~\cite{liu2018fine} identifies and prunes the backdoor neurons. 
Neural attention distillation (NAD)~\cite{li2021neural} first fine-tunes the backdoored network on a clean subset and uses the obtained model as the teacher model. Then NAD utilized this teacher network to guide the fine-tuning of the original backdoored network on the same clean subset to weaken the backdoor neurons. Adversarial neuron pruning (ANP)~\cite{wu2021adversarial} leveraged adversarial neuron perturbation to identify and prune the backdoor neurons. The authors in ~\cite{li2023reconstructive, zeng2021adversarial} leveraged the unlearning method to identify and prune the backdoor neurons.

While these defenses have been proven to be successful on full-precision backdoored models, they are less effective on conditioned backdoor attacks, especially quantization-conditioned backdoor attacks~\cite{li2024nearest}. ~\cite{li2024nearest} presents the first defense against quantization-conditioned backdoors and points out that the nearest rounding operation pushes the dormant backdoor to activation in the quantized models. Based on this observation,~\cite{li2024nearest} proposed a non-nearest quantization rounding strategy to counteract backdoor effects.~\cite{li2024purifying} figured out that the activation distribution of both benign and poisoned samples on the backdoor neuron has a notable change after quantization whereas this phenomenon does not exist in the benign neuron. Based on this~\cite{li2024purifying} purified the backdoor-exposed quantized model by aligning its layer-wise activation with its full-precision version. 

\begin{openquestionbox}
\begin{itemize}
\it 
\item To what extent do conditioned backdoor attacks maintain their effectiveness when transferred across different compression techniques in TinyML models?
\item What is the feasibility of implementing on-device defenses against backdoor attacks, such as trigger detection, on resource-constrained TinyML devices? 
\item What are the costs and challenges associated with deploying these defenses for TinyML applications?
\end{itemize}
\end{openquestionbox}

\subsection{Model Inversion Attacks}
\textbf{Overview:} Model inversion attacks (MIA) are a type of privacy breach where an adversary aims to reconstruct the training data used to train a model by exploiting access to the model itself. These attacks can reveal sensitive information by leveraging the model's outputs or internal representations to infer and approximate the original training data~\cite{yuan2023pseudo, yuan2022secretgen}. MIA is particularly concerning for models trained on private or sensitive datasets, such as medical records or personal images.

Early studies directly searched image pixels to perform MIAs, which were limited to shallow networks and low-resolution images. Recent MIAs mostly leverage generative adversarial networks (GAN) to perform MIA on deeper networks and reconstruct high-resolution images~\cite{zhang2020secret, chen2021knowledge, an2022mirror, wang2021variational}. In this review, we mainly focus on GAN-based MIAs as they are the most popular. In a GAN-based Model Inversion Attack (MIA), a generator produces images from latent vectors which are optimized by a discriminator to enhance their resemblance to the training data of the target model. Pre-trained GANs on extensive public datasets are commonly employed in this approach to bypass the substantial computational costs associated with training a GAN from scratch. The primary objective of a GAN-based MIA is to identify the appropriate latent vector that can effectively reconstruct the training data used by the target model. GAN-based MIA can be categorized into a white-box and a black-box setup based on the attacker's capability and knowledge. 

\textbf{Attack Description:} In white-box MIA, attackers have full access to the weights and outputs of the target model, allowing them to infer the latent vectors by analyzing gradients or reconstructing data via optimization techniques. Early work used the Cross-Entropy (CE) loss to do this task~\cite{yuan2022secretgen, zhang2020secret, chen2021knowledge, an2022mirror, wang2021variational}. To mitigate the gradient vanishing problem associated with the CE loss, ~\cite{struppek2022ppa} proposed to employ Poincare loss and~\cite{yuan2023pseudo} proposed to use max-margin loss. Moreover, ~\cite{nguyen2023re} suggested that the main objective of MIA is to reconstruct images that closely align with a target class while CE loss combines this main objective with a second one that makes the reconstructed image deviate from non-target classes. Therefore, ~\cite{nguyen2023re} developed the negative log-likelihood loss to address the main objective only. Another line of white-box MIA explores the impacts of regularization terms which can provide useful prior knowledge~\cite{qi2023model, nguyen2023re}. ~\cite{struppek24smoothing} studies the impact of label smoothing on MIA.

In contrast, in black-box MIAs, attackers can only access the hard or soft labels and use query-based attacks to infer sensitive input information. A straightforward solution towards this is that attackers first randomly generate and select latent vectors which lead to the images predicted as the target label, and then update the selected latent vector only via decimator loss~\cite{yuan2022secretgen}. ~\cite{kahla2022label} proposed to estimate the gradients and use it to find the optimal latent vector. ~\cite{nguyen2024label} transform the black-box setup to a white-box scenario by leveraging multiple surrogate models. With the increasing power of the diffusion model~\cite{ho2020denoising}, ~\cite{liu2024unstoppable} first use a conditional diffusion model to replace the GAN in MIA. Because these attacks may allow attackers to recover sensitive training data, these attacks receive a medium CVSS score of 4.2.

\textbf{Countermeasure Overview:} To defend against MIA, ~\cite{wang2021improving} proposed reducing the mutual information between the input and output of the target model by modifying the loss function during the training procedure, however at the cost of model performance. ~\cite{peng2022bilateral} proposed a bilateral optimization method that minimizes the dependency between inputs and layer outputs while maximizing the dependency between layer outputs and the model output during the training procedure. ~\cite{gong2023gan} proposed to fine-tune the target model on the fake private samples, which are generated by the target model owner by applying MIA on its own model, to mislead attackers into producing incorrect private data reconstruction.

\begin{openquestionbox}
\begin{itemize}
\it 
\item What theoretical frameworks can be developed to comprehensively analyze a neural network's robustness against MIAs in TinyML systems?
\item How can TinyML network owners effectively assess their networks' resilience to MIAs without relying on resource-intensive empirical testing across various MIA techniques? 
\item What alternative assessment methodologies can be developed specifically for TinyML?
\end{itemize}

\end{openquestionbox}

\subsection{Machine Learning Countermeasure Viability for TinyML}
Through this section, we have covered several possible countermeasures that aim to mitigate attacks targeting the embedded ML model. Here, we do an analysis of the viability of these controls in the context of TinyML. Through this analysis, we find that training-based defenses are preferred because they can be implemented prior to model deployment and do not deplete on-board power. Many pre-processing based defenses require additional power, memory, and compute resources, meaning they may not be viable for TinyML use cases.

\begin{enumerate}[noitemsep]
    \item \textbf{Adversarial Training (Adversarial Examples):} Costa and Pinto's findings suggest that adversarial training methods are as effective in quantized TinyML models as they are in their full-precision counterparts ~\cite{Costa2024DavidEdge}. In addition, these defenses do not incur any additional computational overhead, making them a viable option for defending TinyML models.
    \item \textbf{Feature Squeezing (Adversarial Examples):} Feature squeezing makes use of two operations to detect adversarial examples: reducing the color bit depth and spatial smoothing ~\cite{feat_squeeze}. While these operations are inexpensive on traditional platforms ~\cite{feat_squeeze}, the authors of David and Goliath comment that this cost is non-trivial on devices without a DSP or FPU ~\cite{Costa2024DavidEdge}. More research should be done to benchmark the power and memory costs associated with using this defense on TinyML devices.
    \item \textbf{Pixel Defend (Adversarial Examples):} Because Pixel Defend relies on another neural network to denoise adversarial examples, this defense cannot be used on TinyML class devices ~\cite{pix_def}~\cite{Costa2024DavidEdge}.
    \item \textbf{Query Monitoring (Model Extraction):} While proven to be extremely effective at thwarting model extraction attacks, query monitoring mechanisms may present computational overhead that is unpalatable for TinyML class devices. Methods such as PRADA ~\cite{PRADA} make use of statistical methods to analyze the distribution of API queries and detect patterns that deviate from the patterns of legitimate usage. More research should be done to determine if methods similar to PRADA are able to be implemented on TinyML devices.
    \item \textbf{Query Throttling (Model Extraction):} Limiting the number of queries a TinyML device processes in a unit of time will not lead to resource strain, however, it could hamper the utility of the device in some use cases. In addition, this defense simply makes it take longer to extract the model, making it a relatively weak defense. This countermeasure should be used complementary to other more robust model extraction defenses.
    \item \textbf{Pruning (Backdoor):} Identifying and pruning backdoors is a preferred approach for TinyML, as these steps can be taken before deploying the ML model on the device. Additionally, this method has been proven to be effective in enhancing security while maintaining model performance.
    \item \textbf{Trigger Disabling (Backdoor):} Identifying and disabling the triggers of a backdoor attack must be done during the inference process. This approach demands additional on-device computational power and memory resources, making it less suitable for TinyML, where resources are typically limited.
    \item \textbf{Minimize I/O Relationship (Model Inversion):} The defenses explored for model inversion attacks are implemented during the training phase, allowing them to be applied to TinyML models before they are deployed on the device. As a result, these approaches are well-suited for TinyML, as they do not impose additional computational burdens during inference.
    
\end{enumerate}

\section{Future Directions}
\label{sec:future-directions}

Our analysis of TinyML security challenges has revealed several critical areas that require further research and development, which we discuss below. These research directions address the unique security challenges posed by TinyML systems, balancing the need for robust protection with the severe resource limitations of these devices. Working on these guidelines will help create more secure, transparent, and responsible TinyML ecosystems that can be safely deployed across a wide range of applications.

\subsection{Resource-Efficient Security for TinyML}

During our analysis, we made several observations about hardware, software, and ML model countermeasures for TinyML. We found that the most robust and commonly used countermeasures for SCAs and FIAs are too expensive for TinyML devices in terms of die area and computational overhead. In addition, many of the built-in countermeasures on commodity MCUs do not offer much robustness to the attacks we covered. For software challenges, we found that robust communication channel protections such as TLS may not be viable to implement on deep-embedded MCUs running ML inference. The low-power communication protocols frequently used on these devices have some built-in security features, but often have well-known vulnerabilities. Finally, for ML model challenges, the current literature suggests that training-based defenses may be the best way to defend TinyML devices against attacks as they do not incur the additional overhead that pre-processing-based defenses do. However, it is still not well understood how attacks such as model extraction attacks, backdoor attacks, and MIAs affect TinyML models.

Given these observations, we recommend the following research directions. TinyML devices intrinsically have more resource scarcity than many other MCU use cases due to the introduction of a machine learning model~\cite{micronets}. Traditional protections against attacks on these devices' hardware and software components will need to compete for resources or consume additional die areas, which may be impossible to implement in some cases. If not, their implementation may lead to unpalatable latency and power consumption. The first step in moving toward more secure TinyML deployments is understanding how security measures, the ML model, and other necessary applications interact with the limited resources. As such, more work must be done to benchmark these interactions to better understand resource allocation and trade-offs. Furthermore, more work should be done to validate and fully understand the robustness of TinyML models to the attacks covered in this paper and the costs of their relevant countermeasures. With these metrics, we will be able to identify countermeasures that must be redesigned or replaced to be more resource efficient for use in TinyML deployments.

\subsection{Enhancing Transparency and Auditability}

In recent years, there has been a growing demand for transparency, auditability, and responsibility in embedded AI sensing systems due to the diversity of device platforms, use cases, and application-specific requirements. Significant effort is needed to address this requirement in the development of datasheets for TinyML-based ML sensors~\cite{datasheets, Warden2022MachineSensors, gebru2021datasheets}. 

The aim of the datasheets would be to transparently capture the distinct attributes of these systems, including hardware specifications, characteristics of the ML model and data set, end-to-end performance metrics and environmental impacts. Our proposed research directions would complement this effort and provide a system-level understanding of TinyML security challenges and countermeasures, establishing a crucial framework for security assessment across various TinyML systems. This work would not only justify but also lay the groundwork for integrating ML model security components into the proposed datasheets. We can enhance the privacy and security section of these datasheets by incorporating our findings on hardware vulnerabilities, software challenges, and defenses to the ML model. This could include detailed information on potential risks such as unauthorized data access, device authentication vulnerabilities, and resilience against cyberattacks such as spoofing or data manipulation. In addition, it could outline specific countermeasures implemented, such as adversarial training, tamper detection mechanisms, and encryption schema.

\subsection{Responsible TinyML Development and Deployment}

Incorporating our research into a sensor datasheet will provide developers, researchers, and end-users with a useful tool. This integration will facilitate informed decision-making regarding the security implications of implementing TinyML systems, offer insights into balancing performance and security, and support adherence to shifting regulations. Such a comprehensive methodology for documenting and appraising TinyML sensors is essential to encourage responsible TinyML development and deployment across various use cases.

Moreover, despite the growing attention to AI safety in larger-scale ML systems, these concerns are often overlooked or underestimated in the context of TinyML~\cite{stewart2024materiality}. The unique characteristics of TinyML---such as resource constraints, deployment in diverse and often unsupervised environments, and integration into everyday objects---present distinct challenges that are not adequately addressed by current AI safety frameworks. For example, the limited computational capacity of TinyML devices can restrict the implementation of robust fairness or bias mitigation techniques commonly used in larger systems. The distributed nature of TinyML deployments also complicates oversight and governance, making it challenging to ensure consistent ethical standards across all devices.

\section{Conclusion}
\label{sec:conclusion}

In this paper, we have presented a comprehensive analysis of the security landscape for TinyML devices, offering a taxonomy that distinguishes between IoT, EdgeML, and TinyML systems. Our analysis highlights the unique security challenges posed by TinyML devices and the necessity for tailored defense strategies. We developed a threat model specific to TinyML devices, exploring relevant attacks and their conventional defenses. Table \ref{tab:CVSS} provides a synthesis of these attacks, including their severity ratings according to CVSS, along with the vector strings used for each assessment. Our examination of attack vectors and their respective countermeasures revealed that many off-the-shelf security solutions are either impractical or insufficiently robust for TinyML devices due to their severe resource constraints. Furthermore, we identified significant knowledge gaps in understanding how classical machine learning attacks translate to the TinyML domain, emphasizing the need for further research in this area. Throughout our analysis, we have highlighted open research questions in each section, aiming to guide future work in TinyML security. These questions span hardware, software, and ML model security, addressing the multifaceted nature of TinyML vulnerabilities. We have also proposed key directions for future research, including the development of resource-efficient security measures, enhancing transparency and auditability in TinyML systems, and promoting responsible development and deployment practices. On the whole, our work provides a foundation for researchers, developers, and practitioners to build upon, driving innovation in securing these resource-constrained yet powerful devices. 

\section*{Acknowledgments}

The authors would like to thank Matthew Stewart, Professor Jianfei Yang, and Souvik Kundu for the insights they provided in scoping and framing the structure of this paper and the feedback they gave throughout the writing process.
\bibliographystyle{IEEEtran}
\bibliography{references}

\begin{IEEEbiography}[{\includegraphics[width=1in,height=1.25in,clip]{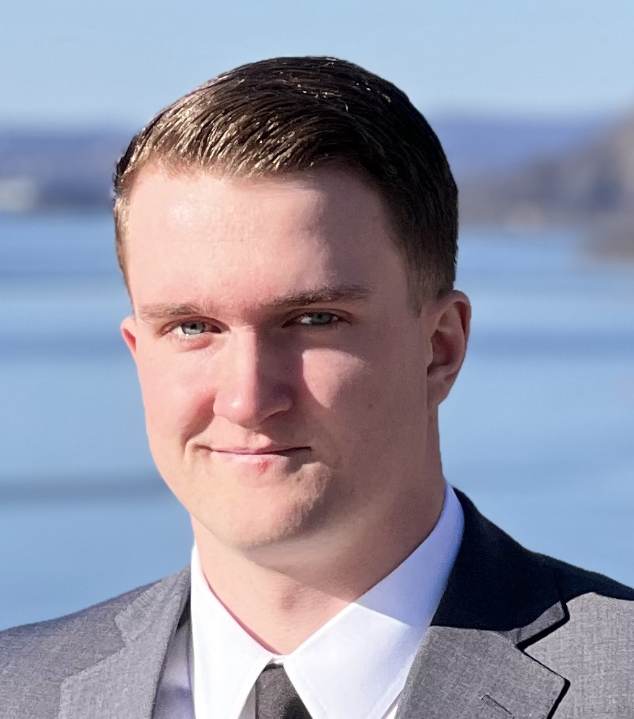}}]{Jacob Huckelberry}
received his B.Sc. in Computer Science from the United States Military Academy in 2023 and is currently a second year student pursuing his M.Sc. in Data Science at Harvard University. Concurrently, Jacob is a Cyber Officer in the U.S. Army and a Draper Laboratory Scholar in their Cyber AI Tools group. His research interests lie at the intersection of machine learning, embedded systems, and cybersecurity.
\end{IEEEbiography}

\begin{IEEEbiography}[{\includegraphics[width=1in,height=1.25in]{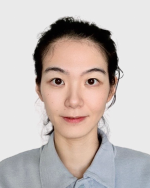}}]{Yuke Zhang}
received the B.Eng. degree from the Beijing University of Posts and Telecommunications, Beijing, China, and the M.A.Sc. from Dalhousie University, Halifax, NS, Canada. She is currently pursuing the Ph.D. degree in electrical and computer engineering with the University of Southern California, Los Angeles, CA, USA. Her research interests include privacy-preserving machine learning and hardware security.
\end{IEEEbiography}

\begin{IEEEbiography}[{\includegraphics[width=1in,height=1.25in,clip]{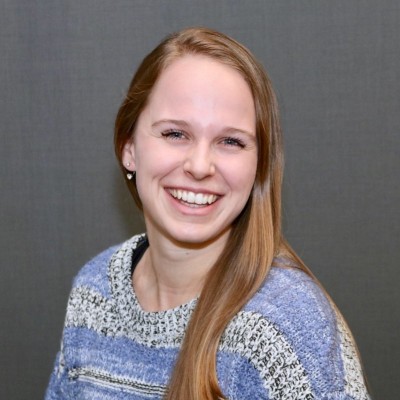}}]{Allison Sansone}
is the Cyber Software and AI Group Leader and a Principal Cyber Software Engineer at Draper Labs in Cambridge, MA. Allison’s expertise lies at the intersection of software reverse engineering and vulnerability analysis, cyber software tool development, and applying machine learning solutions to automate manual cyber processes. Prior to Draper, Allison was a Lead Cyber Software Engineer at MITRE Corporation in Bedford, MA where she performed research in cyber deception, malware analysis, and reverse engineering automation. Allison received her BA in Mathematics and Computer Science from the College of the Holy Cross and her MBA from Worcester Polytechnic Institute.
\end{IEEEbiography}

\begin{IEEEbiography}[{\includegraphics[width=1in,height=1.25in,clip]{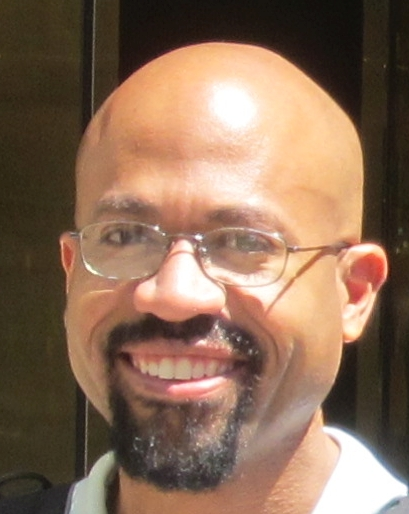}}]{James Mickens}
is a Gordon McKay Professor of Computer Science at Harvard University. His research focuses on the performance and security of large-scale online services. He leverages techniques from isolated software architectures, applied cryptography, and secure hardware design. Prior to becoming a professor at Harvard, Dr. Mickens spent seven years at Microsoft Research, working in the Distributed Systems group and collaborating with various product teams from Azure and Windows; he was also a visiting professor at MIT's Parallel and Distributed Operating Systems group. At Harvard, he serves on the Board of Directors for the Berkman Klein Center for Internet \& Society.
\end{IEEEbiography}

\begin{IEEEbiography}[{\includegraphics[width=1in,height=1.25in,clip]{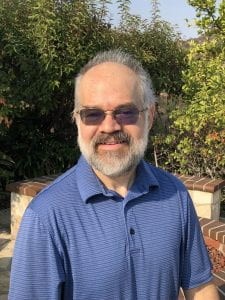}}]{Dr. Peter A. Beerel}
is the Associate Chair of Graduate Programs in the ECE Department at USC and a Principal Distinguished Scientist and Research Director at the Information Sciences Institute. He co-founded TimeLess Design Automation in 2008, which was sold to Fulcrum Microsystems in 2010 then acquired by Intel in 2011. He has (co)authored 16 US patents and more than 190 conference and journal papers covering a wide range of topics in VLSI, CAD, and Machine Learning, several of which have won best paper awards. He has been on the program committee of various VLSI conferences and has been associate editor for IEEE TCAD and IEEE Transactions on Circuits and Systems I.
\end{IEEEbiography}

\begin{IEEEbiography}[{\includegraphics[width=1in,height=1.25in,clip]{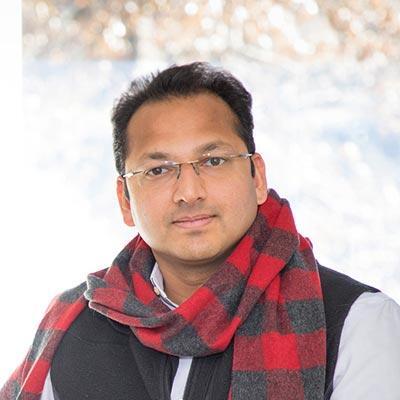}}]{Vijay Janapa Reddi}
expertise lies in computer architecture, machine learning systems, and autonomous agents. His leadership roles in MLCommons and the tinyML Foundation, along with his work in developing the MLPerf benchmarks, provide valuable insights into the practical challenges and opportunities of deploying TinyML in the real-world. Prof. Janapa Reddi's experience in academia and industry collaborations ensures a comprehensive and practical approach to addressing the resource-constrained challenges of securing TinyML systems.
\end{IEEEbiography}

\end{document}